\documentclass[journal]{IEEEtran}
\IEEEoverridecommandlockouts
\usepackage{cite}
\usepackage{booktabs}
\usepackage{multirow}
\usepackage{amsmath,amssymb,amsfonts}
\usepackage[linesnumbered,ruled]{algorithm2e}
\usepackage{graphicx}
\graphicspath{{Fig/}}
\usepackage{textcomp}
\usepackage[dvipsnames]{xcolor}
\usepackage{colortbl}
\usepackage{caption}
\usepackage{url}
\usepackage{float}
\usepackage{lipsum}
\usepackage{cuted}
\usepackage{adjustbox}
\usepackage{mathtools}
\usepackage{flushend}
\usepackage {harpoon}
\usepackage[short]{optidef}
\usepackage[super]{nth}
\usepackage{stmaryrd}

\newcommand{\mysslash}{\mathrel{\mkern-5mu\clipbox{0 0.75ex 0 0}{${\sslash}$}\!}}

\usepackage{hyperref}
\hypersetup{
	colorlinks=true,
	linkcolor=blue,
	citecolor=blue,
	urlcolor=blue
}

\ifCLASSOPTIONcompsoc
\usepackage[caption=false,font=normalsize,labelfont=sf,textfont=sf]{subfig}
\else
\usepackage[caption=false,font=footnotesize]{subfig}
\fi

\begin{document}
\setlength{\abovedisplayskip}{3pt}
\setlength{\belowdisplayskip}{3pt}

\title{A Complex-Coefficient Voltage Control for Virtual Synchronous Generators for Dynamic Enhancement and Power-Voltage Decoupling 
}
\author{Jingzhe~Xu,~\IEEEmembership{Student~Member,~IEEE}, Weihua~Zhou,~\IEEEmembership{Member,~IEEE}, and Behrooz~Bahrani,~\IEEEmembership{Senior~Member,~IEEE}}
%\thanks{This work has been supported by the Monash Grid Innovation Hub and the Australian Renewable Energy Agency (ARENA) under the Advancing Renewable Program (Grant No.: 2020/ARP007).  
%\par The authors are with the Department of Electrical and Computer Systems Engineering, Monash University, 3800 Victoria, Australia.}}
%\markboth{}%
% \markboth{Journal of \LaTeX\ Class Files,~Vol.~17, No.~8, December~2022}%
\maketitle

\begin{abstract} 
As electric power systems evolve towards decarbonization, the transition to inverter-based resources (IBRs) presents challenges to grid stability, necessitating innovative control solutions.
{Virtual synchronous generator (VSG) emerges as a prominent solution. However, conventional VSGs are prone to {instability in strong grids}, slow voltage regulation, and coupled power-voltage response.
To address these issues, this paper introduces an advanced VSG control strategy.} 
A novel analysis of the VSG control dynamics is presented through a second-order closed-loop complex single-input single-output system, employing a vectorized geometrical pole analysis technique for enhanced voltage stability and {dynamics}.
{The proposed comprehensive controller design mitigates issues related to control interacted subsynchronous resonance and $dq \leftrightarrow 3\phi$ transformation-induced voltage-coupled power transients, achieving improved system robustness and simplified control tuning.} Key contributions include {a two-fold design: optimized voltage transition characteristics through direct pole placement and transient power overshoot correction via a compensator.} Validated by simulation and experiments, the findings offer a pragmatic solution for integrating VSG technology into decarbonizing power systems, ensuring reliability and efficiency.

\end{abstract}

% \IEEEpeerreviewmaketitle

\vspace{-0.2cm}

\section{Introduction} \label{sec:Intro}

    % {\color{red}(PS transition brings instability)}\\
    \IEEEPARstart{T}{he} electric power system is shifting towards decarbonization to combat climate change, primarily by adopting inverter-based resources (IBRs) while phasing out fossil-fuel synchronous generators (SGs). 
    This transition, however, deteriorates the grid strength and threatens the security of power systems \cite{kroposki2017achieving}. In response, grid-forming inverter (GFMI) technology has emerged as a promising solution to enhance grid strength \cite{lasseter2019grid}. 

    {A significant aspect of GFMI technology is the virtual synchronous generator (VSG). Extensively researched in academia and industry \cite{d2015virtual}, VSG technology aims to replicate the behavior of conventional SGs, playing a key role in grid synchronization and voltage waveform shaping. {A generic} VSG control scheme features a cascade three-layer primary-voltage-current loop, functioning within a synchronous $dq$-frame. The terminal responses of IBRs are expected to be quick, robust, and resilient, as required by research and legislative frameworks \cite{ramasubramanian2023asking}. However, issues arise with the slow response of the generic VSG controller to disturbances, which impedes VSG integration in strong grids due to its inherent {potential} of instability and volatility\cite{rosso2019robust,wang2020grid}.}

    {A particular challenge is the interaction between the generic VSG controller and the low source impedance, leading to sub-synchronous resonance. This resonance significantly hampers control speeds and exacerbates instability {potentials \cite{shah2023testing,zhao2023closed}. The instability is primarily attributed to slow-responding voltage control and interactive power dynamics \cite{liao2019sub, harnefors2018robust, dokus2021coupling, li2021unified, zhao2022power}.} To address these problems, efforts are commonly put into independently refining the inner voltage-current (VC) and outer primary loops by using the complex transfer function (CTF) \textbf{}\cite{briz2000analysis,harnefors2007modeling} and designing the $2\times2$ controller to decouple real and reactive power ($PQ$), respectively.
    
    Regarding VC loop design, complex coefficients enable two-dimensional flexibility of gain choices while maintaining the symmetry of the CTF. 
    Refs.~\cite{doria2018complex} and \cite{serra2020complex} determine complex poles from the extended Bode plot and the solutions of the high-order polynomial, respectively. Ref.~\cite{quan2020improved} applies the dominant pole-zero cancelation technique by freely choosing complex feedback gains of terminal voltage and current. However, these methods that involve complex gain tuning face challenges in indirect or specialized procedures and computational intensity. Additionally, many of these approaches are primarily designed for off-grid modes, leaving the dynamics of online voltage under-explored.       
    Recently, the real-evaluated CTF control strategies have shifted towards asymmetrical voltage controls, particularly for strong grid applications.
    Ref.~\cite{zhao2022robust} utilizes $q$-axis voltage to dampen $d$-axis current oscillation via active susceptance.
    This scheme does not strictly require coordination that the voltage loop is faster than the power loop. However, it introduces a trade-off, potentially diminishing voltage magnitude controllability due to cross-coupled disturbances.

    {Static $PQ$ decoupling involves using a rotational power frame to match grid impedance angle \cite{Wu2016Unified,li2017improved,He2014Systematic}. To adapt to grid topology variations, event-based adaptive tuning strategies are proposed by \cite{pouresmaeil2021adaptive,mohammed2022online}. Full-state feedback $PQ$ decoupling is also proposed in \cite{chen2023power}. However, these solutions neglect power transients due to static $2\times2$ cross-decoupling gain designs.}     
    To address dynamic $PQ$ coupling, \cite{zhao2022power} proposes a novel approach.
    This method revolves around designing decoupling compensators that invert the dynamics induced by power control interacting with the circuit.
    However, due to the nonlinearity of power transfer, this approach necessitates real-time updates of compensator parameters upon set-point changes. Ref.~\cite{rathnayake2022multivariable} designs the decoupling compensators to reshape the real and reactive power loops to match desired transfer functions with minimized error closely. Despite its adaptability, it inevitably necessitates optimization procedures and is hindered by its complexity.
    {
    Ref.~\cite{Wang2023EffectivePQ} provides an online iterative optimized LQR control for $PQ$ decoupling notable for its adaptivity without prior knowledge of the grid.
    The proposal symmetrizes $PQ$ controllers to regulate frequency and voltage derivative known as complex frequency control\cite{ComplexFreq2022Milano}. 
    However, the real-time optimization poses a computational burden, hindering its practicality.}
    A notable limitation of these strategies is their exclusion of the inner VC loop dynamics.

    \par
    The works mentioned above separately reshape the voltage dynamics by pole-zero manipulation strategies on CTFs and design power decoupling controls by rectifying circuit dynamics or minimizing response errors. 
    Despite the control improvements, nevertheless, the works above
    {do not address the collective solutions of the control interacted subsynchronous resonance of the generic VSG control (vSSCI). Furthermore, the obvious and precise characteristic equation of the grid-connected VC control is yet to be developed}. In addition, the impact of Park-Clark{e} and its inverse transformations ($dq \leftrightarrow 3\phi$)
    identified in \cite{zou2022modeling,zhou2023identification} are neglected.
    To fill these gaps, this paper delineates the explicit and precise voltage behavior of a generic grid-connected VSG by a second-order closed-loop complex single-input single-output (cSISO) system.
    The dominant pole movement is therefore analytically derived.
    The geometrical analysis technique based on mapping the pole onto a complex unit circle successfully justifies the stability margin. {The introduction of complex current feeding gain greatly improves the terminal voltage dynamics. The designed voltage-angle compensator suppresses $dq \leftrightarrow 3\phi$ induced power overshoot. The two-fold design brought dynamics and robustness}
    improvements are validated by simulation and experiments.
    
    \begin{figure*}
        \centering
        \includegraphics[width=0.8\linewidth]{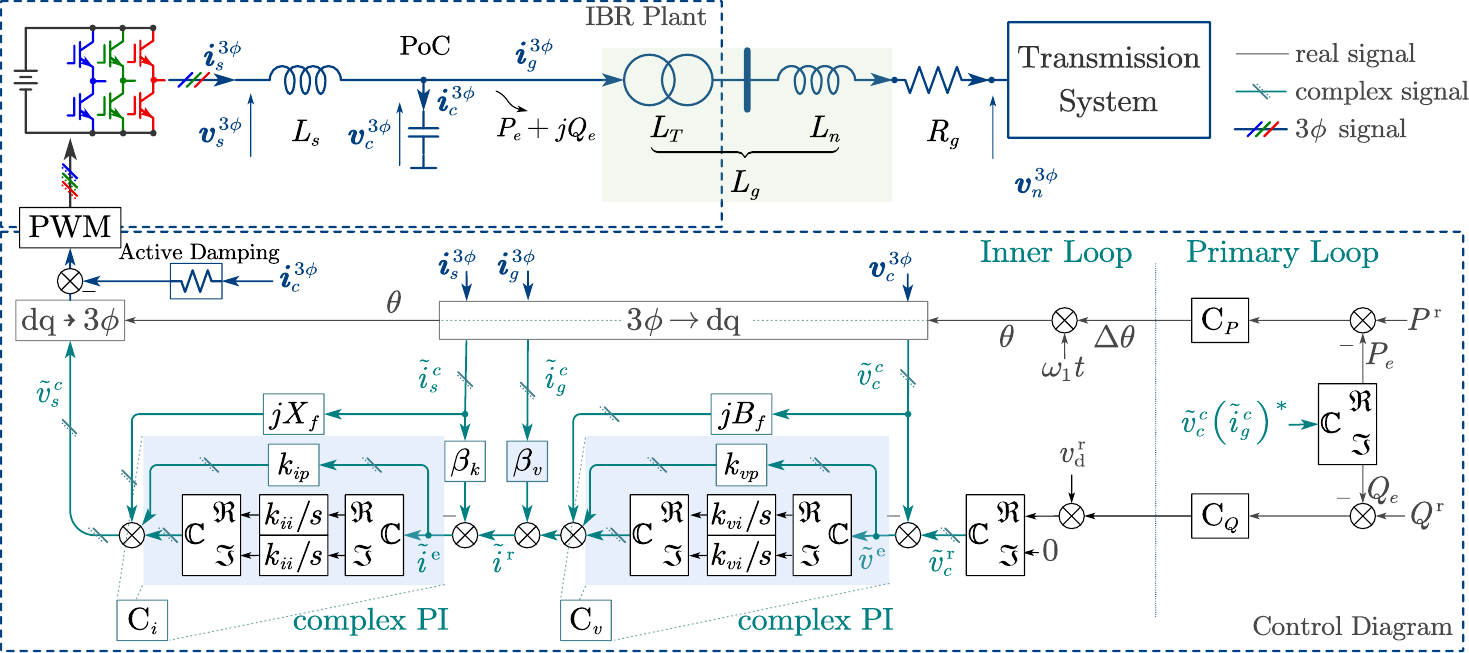}
        \captionsetup{singlelinecheck = false, format= plain, justification=justified, font=footnotesize, labelsep=period}
        \caption{The circuit and control diagram of a VSG in SMIB case. The control loop is formed with mixed real and complex signal flows in Simulink\textsuperscript{\tiny\textregistered}. The detailed time-domain complex PI implementation is exemplified.
        }
        \label{fig:0-SysDiagram}
        % \vspace{-.8cm}
    \end{figure*}

    The contribution of this work can be summarized as follows:
    \begin{enumerate}
        \item {A comprehensive approach to solving the control interacted subsynchronous resonance of a generic VSG coupled with a strong inductive grid via rigorous voltage regulation and transient power compensation; }
        \item A theoretical explanation of the oscillatory {voltage} response of a generic VSG using the vectorized geometrical pole analysis technique of the closed-loop voltage cSISO system;
        \item Development of a complex current feeding gain solution via direct pole placement, achieving sound voltage transition features, e.g. 20~ms rise time and 5\% of overshoot;
        \item Suppression of $dq \leftrightarrow 3\phi$ induced transient {power} overshoot using a simple first-order compensator.
       \end{enumerate}

    {This paper distinguishes itself through its robustness to grid condition changes, straightforward tuning, and minimal deviation from the generic VSG controller.} The rest of this paper is organized as follows. Section~\ref{sec:2} examines a generic VSG in the single-machine-infinite-bus (SMIB) case and investigates the root causes of undesired responses. Section~\ref{sec:3} introduces the pole placement technique and the voltage-angle compensator. Section~\ref{sec:4} presents simulation and experimental validations. Finally, Section~\ref{sec:5} draws the conclusion.
    %\vspace{-.3cm}
    
\section{Undesired VSG Terminal Performance} \label{sec:2}

    In this section, a grid-connected $LC$-filtered VSG is modeled within the $dq$-frame using complex vectors. This approach not only illustrates the intricate dynamics of the system but also sheds light on the undesired voltage and active power terminal responses. Furthermore, the causation of the voltage behavior is investigated via the pole movement of the cSISO voltage transfer function. Finally, the dynamic power fluctuation induced by the voltage perturbation is examined. %\vspace{-.5 cm}

    \subsection{Description of Grid-connected VSG System} \label{sec:2-1}%\vspace{-.1cm}
    Fig.~\ref{fig:0-SysDiagram} depicts the circuit and control diagram of a VSG in a SMIB setup, which represents a high-power (4 MVA) low-switching-frequency (2.5 kHz) solar inverter in a utility-scale IBR plant. The inverter incorporates an $LC$ filter. Its point-of-coupling (PoC) further links the step-up transformers and upstream transmission system, which comprise the equivalent grid impedance. The VSG control scheme comprises a three-layer cascade loop formulated with mixed real and complex signals in the $dq$ domain.

    \begin{figure}
        \centering
        \includegraphics[width=0.9\linewidth]{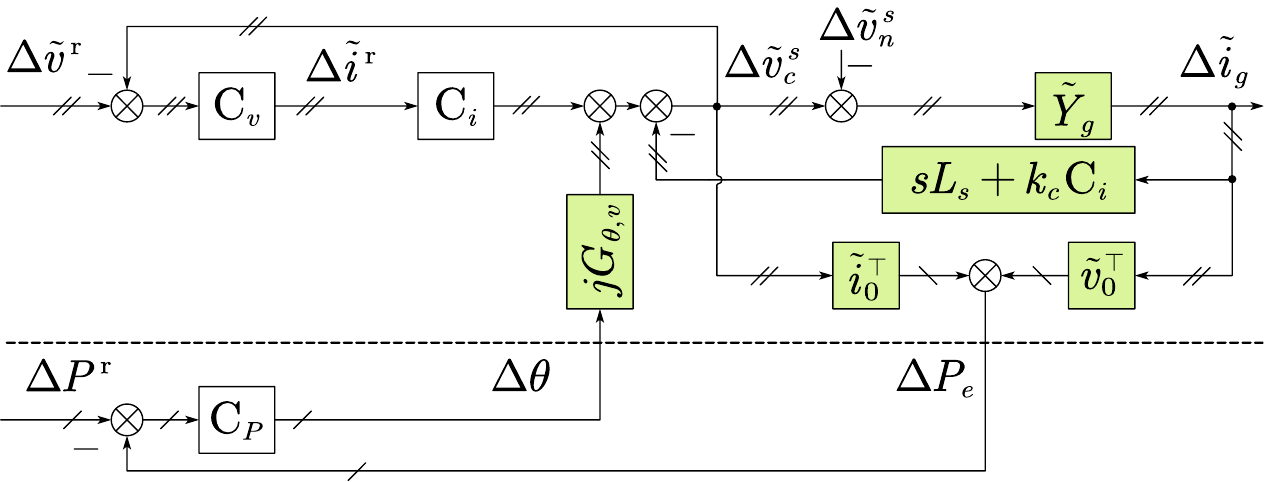} 
        \captionsetup{singlelinecheck = false, format= plain, justification=justified, font=footnotesize, labelsep=period}
        \caption{A closed-loop complex small-signal diagram of a SMIB case VSG.}
        \label{fig:2.1-PowrCL}
        % \vspace{-.3cm}
    \end{figure}
    
     Fig.~\ref{fig:2.1-PowrCL} depicts the closed-loop small-signal model (SSM) block diagram of the system. The outer primary loops are the swing equation for power synchronization and reactive power voltage droop, respectively. The cascade VC loops are symmetrical complex proportional-integral (PI) controls, where the implementation details are exemplified in Fig.~\ref{fig:0-SysDiagram}. The detailed SSM is derived in Appendix \ref{appdx} which excludes high-frequency dynamics.
     {
     One of the dynamics is the $LC$ filter resonance peak, $\omega_1/ \sqrt{X_s B_C}$, in hundreds of Hertz\cite{Liserre2005DesignLCL}. In addition, relevant passive and active resonance suppression schemes are developed in \cite{pan2014OptimizedLCL,Dannehl2011Filter-Based}. Another type of dynamics is the switching caused control delay beyond hundreds of Hertz and can be mitigated by passivity-oriented design \cite{harnefors2015passivity,Wu2021Passivity,Liao2020Passivity}.}
     To focus on voltage regulation stiffness, one of the fundamental features of GFMIs \cite{harnefors2020universal}, the reactive power loop is discarded. 
    Table~\ref{tab:Circuit_and_Controller_Parameters_of_the_VSG} lists the circuit and control parameters of the base case.
    The depicted transformer in Fig.~\ref{fig:0-SysDiagram} comprises two-stage step-ups with approximately summed 15\% impedance at the rated plant VA capacity as per standards \cite{ieeeStd2800,iec60076-5n}. The minimum impedance beyond PoC that an inverter may experience is the first transformer with 4\% impedance at the rated inverter VA base. It sets the lower boundary of $X_g$ in this paper. The definitions of reactance $X=\omega_1 L$ and susceptance $B=\omega_1 C$ are extrapolated in this work for unified expressions.  %\vspace{-.5 cm}

    \begin{table}[]
    	\centering
    	\captionsetup{singlelinecheck = false, format= plain, labelsep=newline, font={sc, footnotesize}, justification=centering}
    	\caption{Circuit and Control Parameters of the Studied VSG}
        %\vspace{-.15cm}
    	\label{tab:Circuit_and_Controller_Parameters_of_the_VSG}
        \begin{tabular}{llc}
        \hline \hline
        Symbol     & Description                                  & Value                      \\ \hline
        $V_\mathrm{dc}$   & DC-link voltage                              & 1500 V                     \\
        $\omega_1$ & Electrical angular frequency base            & $50 \cdot 2\pi$~rad/s      \\
        $V_{ll,1}$ & Three-phase voltage base                     & 690 V                      \\
        $X_{s}\,  (\omega_1 L_s)$     & Filter reactance        & 0.10 p.u.                  \\
        $X_{g}\,(\omega_1L_g)$    & Grid reactance                & 0.30 p.u.                  \\
        $R_{g}$    & Circuit equivalent series resistance (ESR)   & $10^{-3}$ p.u.             \\
        $B_C\, (\omega_1 C_{f})$      & Filter susceptance      & 0.01 p.u.                  \\
        $k_{ip}$   & Current controller proportional gain         & 0.4776 p.u.                \\
        $k_{ii}$   & Current controller integral gain             & 15 p.u./s                  \\
        $k_{vp}$   & Voltage controller proportional gain         & 0 p.u.                     \\
        $k_{vi}$   & Voltage controller integral gain             & 800 p.u./s                 \\
        $\beta_v$  & Voltage loop feed-forward ratio              & 0.5                        \\
        $\beta_k$  & Current feedback ratio                        & 1                          \\
        $H$        & Inertial time constant                       & 1 s                        \\
        $D$        & Angular speed damping ratio                  & 66.67 p.u.                 \\
        $K_q$      & Reactive power droop ratio                   & 0                          \\ \hline\hline
        \end{tabular}
        %\vspace{-.5cm}
    \end{table}

    \subsection{Problem Replication} \label{sec:2-2} %\vspace{-.1cm}
    The conventional VC loop tuning applies symmetry optimum (SO) \cite{yazdani2010voltage}, which roughly maintains the cross-over frequency of the voltage and current PIs in a certain ratio with a desired phase margin. This method sets the VC loop by $\{k_{vi},k_{ip}\}=\{22.1,0.796\}$. The fractional grid current feed-forward gain $\beta_v = 0.5$ helps stabilize the system \cite{ravanji2023impact}. 
    The system is perturbed with 0.05 p.u. voltage set-point step-change at 1~s.
    It results in the voltage magnitude response of 100~ms rise time, 56\% of overshoot, and 2.105~Hz oscillation.
    In addition, the active power is perturbed with 54\% of overshoot as depicted in Fig.~\ref{fig:2.0-VoltResponseComp}(a).
    The SO method-tuned controller shows a weak terminal regulation \cite{d2014automatic}.
    Further exaggeratedly increasing $k_{vi}=800$ as the base case, the voltage rise time is limited to 10 ms, and the overshoot is 32\%. However, the coupled active power exacerbates an exaggerated 370\% of overshoot. The time-domain simulation shows a 30.3~Hz mode as depicted in Fig.~\ref{fig:2.0-VoltResponseComp}(b). Derived from the SSM, the closed-loop frequency responses of the voltage magnitude and active power to the voltage set-point perturbation of these two configurations are shown in Fig.~\ref{fig:2.0-VoltResponseComp}(c). It captures the vSSCI at 2.18~Hz and 30.1~Hz of these two systems, respectively.
    
    In summary, the time-domain simulation shows that the generic VSG controller {can barely} achieve a fast and properly damped response simultaneously. Furthermore, the amplified power volatility risks intolerable hardware damages \cite{ravanji2023comparative}. The root causes of the undesired voltage dynamics and the caused active power coupling are explored in the rest of this section.
    %\vspace{-.3 cm}
    \begin{figure}
        \centering
        \includegraphics[width=1\linewidth]{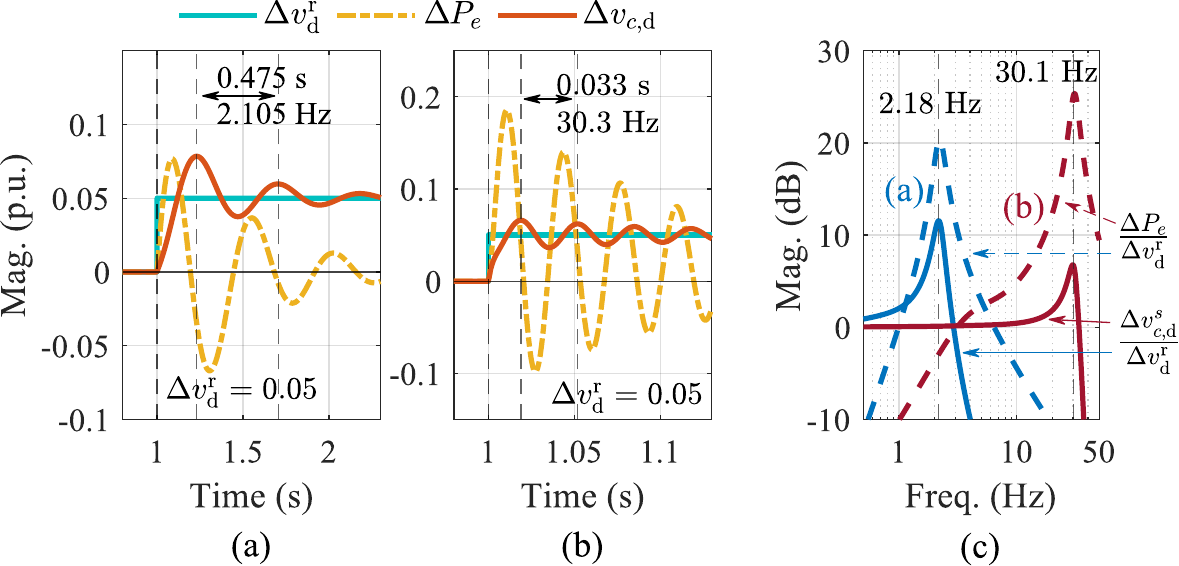}
        \captionsetup{singlelinecheck = false, format= plain, justification=justified, font=footnotesize, labelsep=period}
        \caption{
        Comparison of voltage and active power response to voltage reference perturbation by different tuning methods: (a) The SO $\{k_{vi}, k_{ip}\}=\{22.1,0.796\}$ and (b) the base case with excessive high voltage integral gain $k_{vi} = 800$. (c) The voltage and real power frequency responses to the voltage set-point change with the configurations described in (a) and (b), forming the resonance peak at 2.18~Hz and 30.1~Hz, respectively.
        }
        \label{fig:2.0-VoltResponseComp}
        %\vspace{-.8cm}
    \end{figure}

    \subsection{Explanation of {Axial} Cross-Coupled Voltage Responses} \label{sec:2-3}
    %\vspace{-.3cm}
    {Focusing on the reference perturbation $ \Delta \tilde{v}^{\mathrm{r}}$ in (\ref{eq:5}), the open-loop complex transfer function {of VC loop} is expressed as:}
    \begin{align}
        \label{eq:11a}
        \frac{\Delta \tilde{v}_c}{\Delta \tilde{v}^{\mathrm{e}}_c}=\frac{\Delta \tilde{v}_c}{\Delta \tilde{v}^{\mathrm{r}}-\Delta \tilde{v}_c}=\frac{\mathrm{C}_i\mathrm{C}_v}{L_P+1},
    \end{align}
    with the open-loop plant given by:
     \begin{align}
        \label{eq:11b}
        L_P=\frac{sL_s+ k_c\mathrm{C}_i}{sL_g+R_g+{j}X_g}{=\frac{sL_s+ k_c\mathrm{C}_i}{R_g+(s+{j}\omega_1)L_g}},
    \end{align} 
    as depicted in Fig.~\ref{fig:2-VoltCL}, the block diagram of the complex voltage loop.
    Observing (\ref{eq:11b}), the grid impedance introduces an open-loop complex pole, {$p_\mathrm{ol, plant}= -R_g/L_g - j\omega_1$,} into the composed plant.
    {The lack of complex coefficients in the plant zero by the composed current controller results in cross-coupling and poor damping in $dq$-axis voltage responses.}
    {In detail, a voltage magnitude reference step $\Delta v_{\mathrm{d}}^{\mathrm{r}}$ perturbs PoC $d$-axis voltage $\Delta v_{c,\mathrm{d}}$. It then excites $q$-axis grid current $\Delta i_{g,\mathrm{q}}$ via ${j}X_g$. $\Delta i_{g,\mathrm{q}}$ further perturbs PoC $q$-axis voltage $\Delta v_{c,\mathrm{q}}$ via the composed current controller. While $v_q^{\mathrm{r}} = 0$ holds constantly, the controller $\mathrm{C}_v$ is to restore the $q$-axis voltage discrepancy. It further triggers the perturbations of $\Delta v_{c,\mathrm{q}}$, $d$-axis $\Delta i_{g,\mathrm{d}}$, and $\Delta v_{c,\mathrm{d}}$ sequentially by the same mechanism.}
    
    \par To evaluate dynamic performance, the voltage closed-loop CTF is derived from (\ref{eq:11a}) as:
    \begin{align}
        \label{eq:12_}
         {\tilde{G}_{v,\mathrm{cl}}(s)} = \frac{\Delta \tilde{v}_c}{\Delta \tilde{v}^{\mathrm{r}}}\mid_{\mathrm{cl}}=\frac{b_1s+b_0}{a_2s^2+a_1s+a_0},
    \end{align}
    where
    \begin{align}
        \nonumber
        &a_0=b_0={j}X_gk_{ip}k_{vi},    &b_1=L_gk_{ip}k_{vi},    \\
        % \label{eq:12_b}
        \nonumber
        &a_1=k_ck_{ip}+L_gk_{ip}k_{vi}+{j}X_g,    &a_2=L_g+L_s, 
    \end{align}
    with simplifications: $R_g = 0$, $\mathrm{C}_v = k_{vi}/s$, and $\mathrm{C}_i = k_{ip}$ \cite{harnefors2018robust}. 
    The control canonical form of (\ref{eq:12_}) with $\Delta \tilde{v}^{\mathrm{r}}$ as the input, $\Delta \tilde{v}_c$ as the output, and two arbitrary states, the realized state-space system $\left( \boldsymbol{A}_{v},\boldsymbol{B}_{v},\boldsymbol{C}_{v}, {D}_{v} \right) $ is defined as \cite{bishop2011modern}
    \begin{align}
        \boldsymbol{A}_{v}&=\left[ \begin{matrix}
        	-\frac{a_1}{a_2}&		-\frac{a_0}{a_2}\\
        	1&		0\\
        \end{matrix} \right] , \,\boldsymbol{B}_{v}=\left[ \begin{array}{c}
        	1\\
        	0\\
        \end{array} \right] ,\,            % \\
        \boldsymbol{C}_{v}=\left[ \begin{matrix}
    	\frac{b_1}{a_2}&		\frac{b_0}{a_2}\\
    \end{matrix} \right] ,
    \end{align}
    and {$D_{v}=0$. % The eigenvalues of $\boldsymbol{A}_{v}$ or
    The complex poles of the realized system are}
    \begin{align}
        \label{eq:v_poles}
        \lambda _{1,2}&=-\frac{a_1\pm \sqrt{{a_1}^2-4a_0a_2}}{2a_2},
    \end{align}
    which could be reformulated as 
       \begin{align}
       \label{eq:v_poles, alt}
        \lambda _{1,2}&=-M\underset{\mathrm{Angle \, dominant}}{\underbrace{\left[ e^{j\phi}\pm \sqrt{e^{-j2\phi}+j\mu} \right] }},
    \end{align} 
     %\vspace{-0.2cm}
    where
    \begin{align}
        % \nonumer
        \label{eq:M, mu}
        % \begin{matrix}
        	M=\left| \frac{a_1}{2a_2} \right|, \,\, \mu =4\frac{k_c-L_sk_{vi}}{|a_1|^2/(X_g k_{ip})}, \,\, \angle \phi = \mathrm{arg} (a_1).
        % \end{matrix}
    \end{align}
   %\vspace{-0.01cm}
    Equation~(\ref{eq:v_poles}) scales down the poles on a complex unit circle by the magnitude of $M$, where $e^{j\phi}$ and ${e^{-j2\phi}}$ are unity vectors. Fig.~\ref{fig:3.0-pole_moving_vector_group} depicts the angle dominant part of (\ref{eq:v_poles, alt}).
    It also manifests the complex movements of the dominant pole, nominated as $\lambda_2$, when $\mu$ deviates from 0.
    If $\mu=0$, the two scaled poles, $\lambda_{1,2}/M = -(e^{j\phi}\pm e^{-j\phi}) = \{-2 \cos{\phi}, -j2 \sin{\phi} \}$, pin on the real and imaginary axes, respectively. The loop is marginally stable.
    If $\mu>0$, $j\mu$ heads to $90^{\circ}$, the magnitude of the square root part is less than 1. Hence, 
    deducting the square root part from $e^{j\phi}$ results in a positive real part, or poles are in the left-half plane.
    If $\mu<0$, conversely, $j\mu$ heads to $-90^{\circ}$, the square root part has an amplified magnitude larger than 1, which consequently leads to forming an unstable pole with a positive real part.
    %\vspace{-.1cm}
        \begin{figure}[!]
            \centering
                \includegraphics[width=1\linewidth]{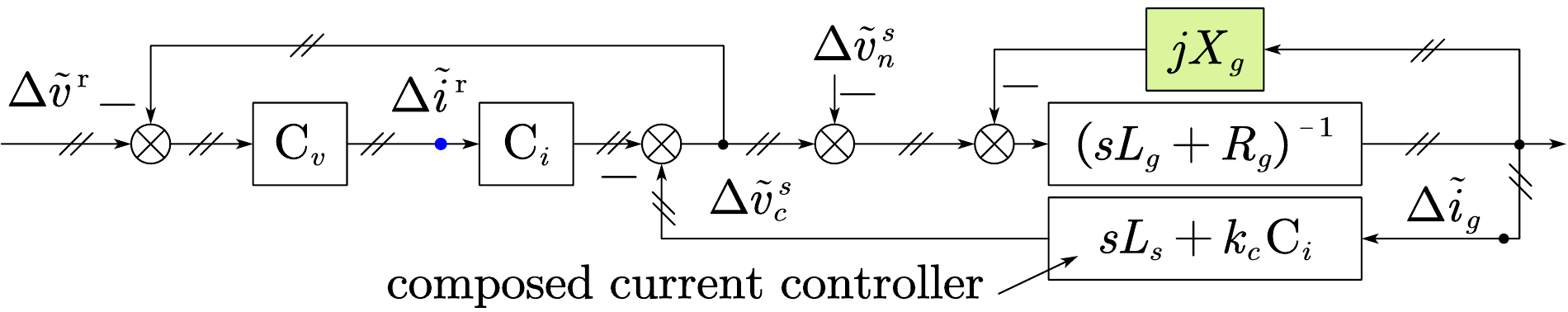}
                \captionsetup{singlelinecheck = false, format= plain, justification=justified, font=footnotesize, labelsep=period}
                \caption{Complex voltage loop of VSG with grid dynamics. The imaginary part of grid admittance is shaded in green.}
                \label{fig:2-VoltCL}
                %\vspace{-.8cm}
        \end{figure}
        \begin{figure}[!]
            \centering
                \includegraphics[width=1\linewidth]{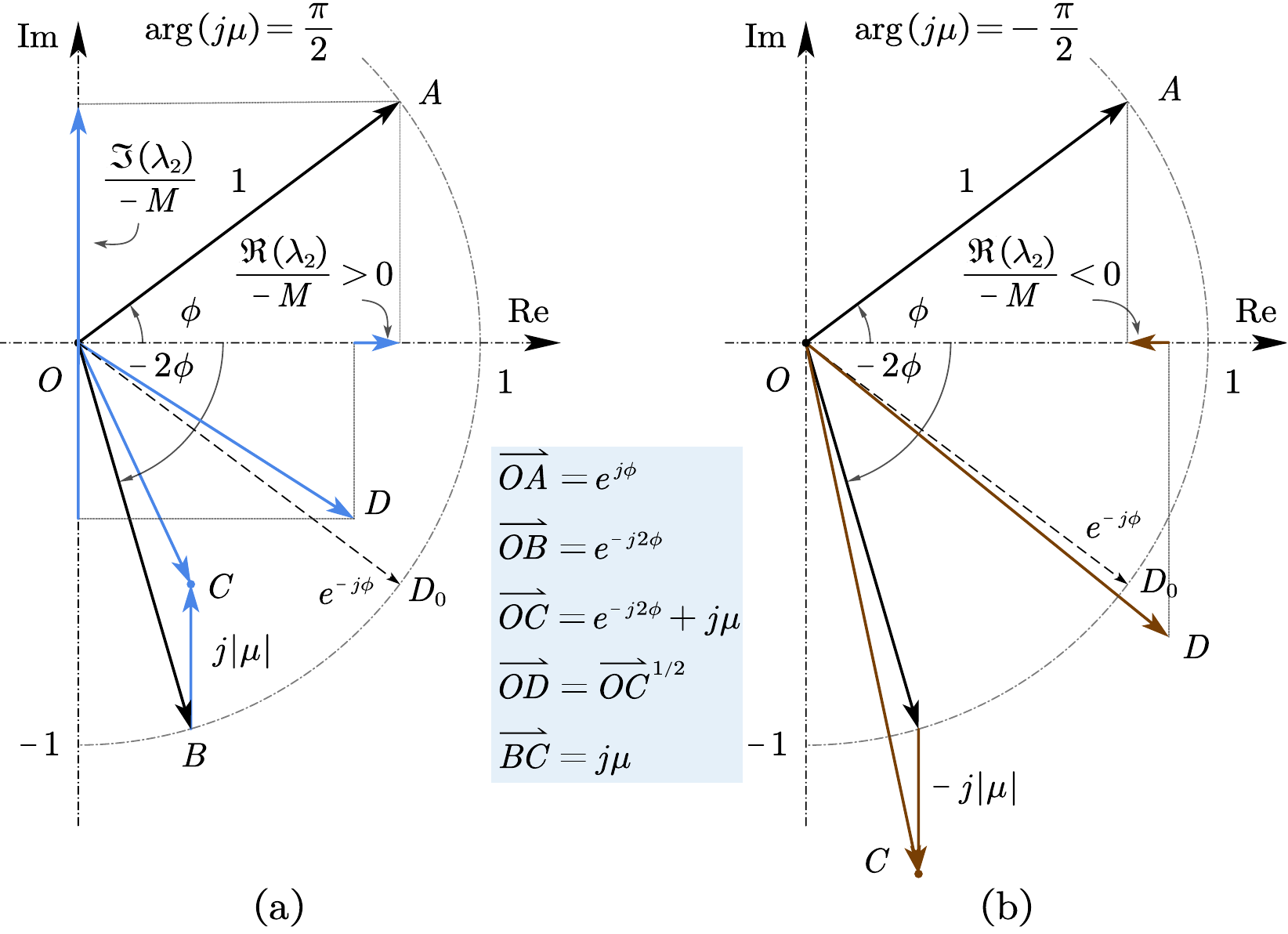}
                \captionsetup{singlelinecheck = false, format= plain, justification=justified, font=footnotesize, labelsep=period}
                \caption{Illustration of the angle dominant part of poles in the complex plane for stability margin identification.
                Points $C, D$ move from $B, D_0$ respectively when $\mu$ deviates from 0:
                (a) The stable dominant pole case with a positive $\mu$, (b) The unstable dominant pole case with negative $\mu$.}
                \label{fig:3.0-pole_moving_vector_group}
                %\vspace{-.8cm}
        \end{figure}
    In the parameter set of the base case, the instability zone, $\mu < 0$, yields $k_c < 0.382$ or equivalently $\beta_v > 0.618$.     
    Fig.~\ref{fig:3.3 time_mu_stability} simulates a scenario where the perturbed voltage response transits from unstable to stable by increasing $k_c$ from 0.35 to 0.40. 
    \begin{figure}[!]
            \centering
                {\includegraphics[width=1\linewidth]{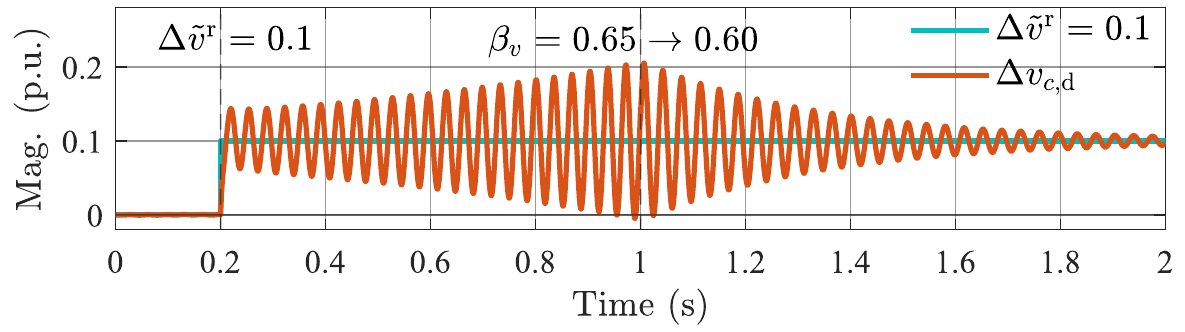}}
                \captionsetup{singlelinecheck = false, format= plain, justification=justified, font=footnotesize, labelsep=period}
                \caption{The PoC $d$-axis voltage transits from unstable to stable by reducing $\beta_v$ from 0.65 to 0.60 at 1 s or equivalently increasing $k_c$ from 0.35 to 0.40 to let $\mu$ regain positive. The voltage reference increases 0.1 p.u. at 0.2 s.}
                \label{fig:3.3 time_mu_stability}
    \end{figure}

    \begin{figure}[!]
            \centering
                {\includegraphics[width=1\linewidth]{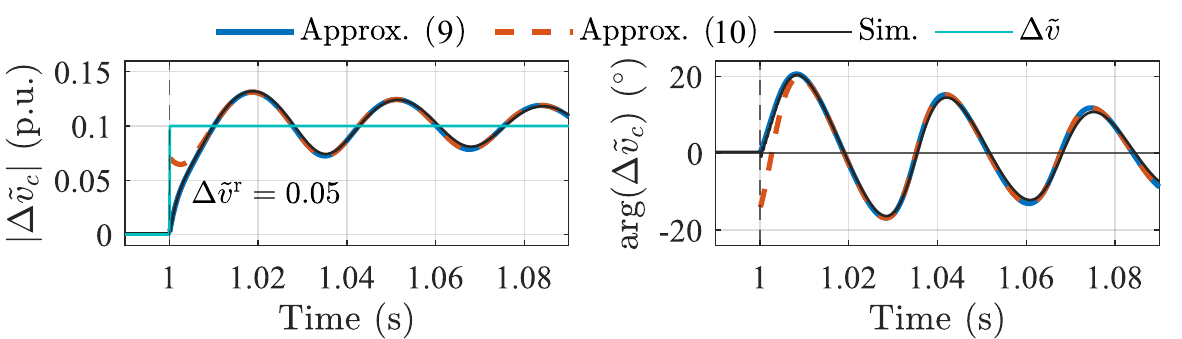}}
                \captionsetup{singlelinecheck = false, format= plain, justification=justified, font=footnotesize, labelsep=period}
                \caption{The complex voltage perturbation $\Delta \tilde{v}_c$ response to the reference step-change at 1 s with the disabled power loop of the base case.}
                \label{fig:3-TimeStep_Comp_VcCplx_response_BaseCase}
                %\vspace{-.7cm}
    \end{figure}

    The system frequency- and time-domain responses are {\cite{vowles2015smallsignal}}
    \begin{align}
        \label{eq:v_freq} 
        {\tilde{G}_{v,\mathrm{cl}}(s)} &=\boldsymbol{C}_{v}\boldsymbol{V}_{v}\left( s\mathbf{I}-\boldsymbol{\Lambda }_{v} \right) ^{-1}\boldsymbol{V}_{v}^{-1}\boldsymbol{B}_{v}+D_{v},
    \end{align}
    and
    \begin{IEEEeqnarray}{rCl}\label{eq:v_step} 
        {\tilde{G}_{v,\mathrm{cl}}(t)} &=&\boldsymbol{C}_{v}\boldsymbol{V}_{v}\left( e^{\boldsymbol{\Lambda }_{v}t}-\mathbf{I} \right) \left( \boldsymbol{AV}_{v} \right) ^{-1}\boldsymbol{B}_{v}+D_{v}
               \IEEEnonumber\\
        &=&1-\frac{\lambda _1\mathrm{e}^{\lambda _2t}-\lambda _2\mathrm{e}^{\lambda _1t}-\frac{b_1}{a_0}\lambda _1\lambda _2\left( \mathrm{e}^{\lambda _1t}-\mathrm{e}^{\lambda _2t} \right)}{\lambda _1-\lambda _2},\IEEEeqnarraynumspace
    \end{IEEEeqnarray}
    respectively, where the corresponding left eigenvectors $\boldsymbol{V}_{v}$  and diagonalized eigenvalues $\boldsymbol{\Lambda}_{v}$ are 
    \begin{align}
        \nonumber
        &\boldsymbol{V}_{v}=\left[ \begin{matrix}
        	\lambda _1&		\lambda _2\\
        	1&		1\\
        \end{matrix} \right] , 
        &\boldsymbol{\Lambda}_{v} = \mathrm{diag}(\lambda_1, \lambda_2).
    \end{align}
    
    The roots of a complex-valued CTF are not conjugate \cite{briz2000analysis}.
    Therefore, the real part of the stable dominant pole $\lambda _2$ is closer to the origin, i.e. $\Re(\lambda _2)=\max \Re \left\{ \lambda _{1,2} \right\} $, which varies the operator $\pm$ in (\ref{eq:v_poles},\ref{eq:v_poles, alt}) and (\ref{eq:pole-placement}) to $-$.
    Eq. (\ref{eq:v_step}) is simplified to
    \begin{align}
        \label{eq:v_step_approx} 
        {\tilde{G}_{v,\mathrm{cl}}(t)}  
        % 1-{\mathrm{e}^{\lambda _2t}}\frac{1-\lambda _2 {j} / \omega_1 }{1-\lambda _2/\lambda _1}
        \approx 1-{\mathrm{e}^{\lambda _2t}}({1-{j} \lambda _2  / \omega_1 }),
    \end{align}
    as the magnitude and angle estimator after minor poles decay provided $|\lambda_1| \gg |\lambda_2|$, where ${j} \omega_1 = \small{{b_1}/{b_0}} $ is the transmission zero. 
    Fig.~\ref{fig:3-TimeStep_Comp_VcCplx_response_BaseCase} compares the magnitude and angle response of $\Delta \tilde{v}_c^s$ to a reference step-change of the base case. Both (\ref{eq:v_step}) and (\ref{eq:v_step_approx}) replicate the simulation responses.
    \par In summary, the inductive grid causes the voltage oscillatory behavior of the generic grid-connected VSG. The proposed geometrical pole method identifies the root cause analytically. The derived vector $j\mu$ sketches the stability margin. 
    % \vspace{-.5cm}
    
    \subsection{Root Cause Analysis of Power-Voltage Coupling and Power Angle Stability}  \label{sec:2-4}   %\vspace{-1mm}
      
    \par The voltage-induced power fluctuation in VSGs is decomposed into the static and dynamic circuit factors. In a purely inductive grid, power transfer is conventionally analyzed using the quasi-steady-state equation $ P= V_s V_r \sin(\delta) /X  $ and its linearized form:
        \begin{align}
            \label{eq:12}
            \Delta P=\frac{V_r}{X}\sin \delta _0\Delta V_s+\frac{V_sV_r}{X}\cos \delta _0\Delta \delta ,
        \end{align}
    where $V_s$ and $V_r$ are the voltage magnitude of the sending and receiving end, respectively. $\delta_0$ is the power angle and $X$ is the reactance \cite{kundur2022power}. 
    However, this approach neglects re-synchronization dynamics, which is crucial in VSG control. {The non-zero $q$-axis voltage convergence to zero contributes to additional angle dynamics.}
    
    The re-synchronization dynamic and stability involve both voltage and active power loops through the grid impedance as
    \begin{align}
        \nonumber
        \Delta P_e =& \left<\left( \tilde{i}_{g}^{0}+\tilde{Y}_gv_{c,\mathrm{d}}^{0} \right)  \cdot   \Delta \tilde{v}_{c}^{s} \right> \\
        \nonumber
        =&\left[ i_{g,\mathrm{d}}^{0}+\frac{\left( sL_g+R_g \right) v_{c,\mathrm{d}}^{0}}{\left( sL_g+R_g \right) ^2+X_{g}^{2}} \right] \Delta v_{c,\mathrm{d}}^s\\
        \label{eq:13}
        &+\left[ i_{g,\mathrm{q}}^{0}+\frac{v_{c,\mathrm{d}}^{0} X_g}{\left( sL_g+R_g \right) ^2+X_{g}^{2}} \right] \Delta v_{c,\mathrm{q}}^s,
    \end{align}
    with zero steady-state of $ v_{c,\mathrm{q}}^{0}$. 
    Eq.~(\ref{eq:13}) shows that the cross-coupled $dq$-axes voltage disturbance perturbs the active power,  
    which includes the static set-point related $\left< \tilde{i}_{g}^{0}\cdot \Delta \tilde{v}_c \right> $ and dynamic circuit involved $\left< \tilde{v}_{g}^{0} \tilde{Y} \cdot \Delta \tilde{v}_c \right> $ amplification, respectively. 
   
    \par The control coupling induced by the $dq \leftrightarrow 3\phi$ frame transformations is vital, too. The term ${j}G_{\theta,v}$ in (\ref{eq:5}) encapsulates the transformation's impact on PoC voltage, {primarily deviating $v_{c,\mathrm{q}}$ from 0 by rotating the inverter's synchronous frame. It inherently perturbs active power as in (\ref{eq:13}). This mechanism results in transient power coupling during voltage transitions.}
    In brief, both the circuit and control contribute to the voltage-induced power coupling of a VSG, seeking further solutions, as reported in subsequent sections. %\vspace{-3mm}
    % {\color{red}subsection summary!}

\section{Proposed Control Improvement Strategies}\label{sec:3} %\vspace{-1mm}
        The solution to the {voltage-induced power coupling}, as analyzed in Section~\ref{sec:2-4}, comprises two stages. 
        First, a fast and well-damped voltage controller is needed for grid connection so that the PoC voltage set-point can be quickly and smoothly tracked, minimizing the impact on power re-synchronization. Hence, the complex current feeding {gain} is proposed accordingly. Second, the $dq\leftrightarrow 3\phi$ induced control coupling can be rectified by inserting a reversal compensator. 
        
        Section~\ref{sec:3-1} provides the complex feeding gain solution, ensuring the voltage loop dynamics and robustness based on the root cause found in Section~\ref{sec:2-3}. Section~\ref{sec:3-2} introduces the voltage-angle compensator which only depends on inner VC loop parameters, further providing the implementation proposal. The frequency-domain validation of the designs on vSSCI suppression is presented in Section~\ref{sec:3-3}. %\vspace{-5mm}
    \subsection{Improved VC Loop by Complex Current Feeding Gain} \label{sec:3-1}  %\vspace{-1mm}
        A fast and well-damped voltage controller is desired for the overall performance of VSG.  
        To achieve this, a few well-recognized criteria for good control are set out:
        a transition duration (from 10\% to 95\%) limited within 1 electrical cycle \cite{doria2018complex},
        poles damping ratio more than 0.707 \cite{yazdani2010voltage},
        a post-transition overshoot within 5\%. 
        In addition, the variation has minimum impacts on existing controls and does not deteriorate the functionalities of other loops, e.g., the current loop damper.

        The following parts provide a complex current feeding gain solution by rotating and elasticizing the vector $j\mu$, parameterizing the controller by dynamic optimization and direct pole placement. The robustness is further validated.
        
        \subsubsection{Voltage Dynamic Criteria and Approach}\label{sec:3-1-1}
        To meet design criteria, the voltage dominant pole $\lambda_2$ needs to increase its natural frequency and damping ratio extensively. 
        To achieve this aim, Fig.~\ref{fig:3.0-pole_moving_vector_group}(a) suggests to 
        rotate the vector $\overrightarrow{BC}$ heading to the second quadrant, i.e., $\mathrm{arg}(j{\mu}) \in (\frac{\pi}{2}, {\pi})$, with an extended magnitude.
        Fig.~\ref{fig:3.1_V_voltage_group_proposal} portrays this scenario qualitatively.
        \begin{figure}[!]
            \centering
            \includegraphics[width=1\linewidth]{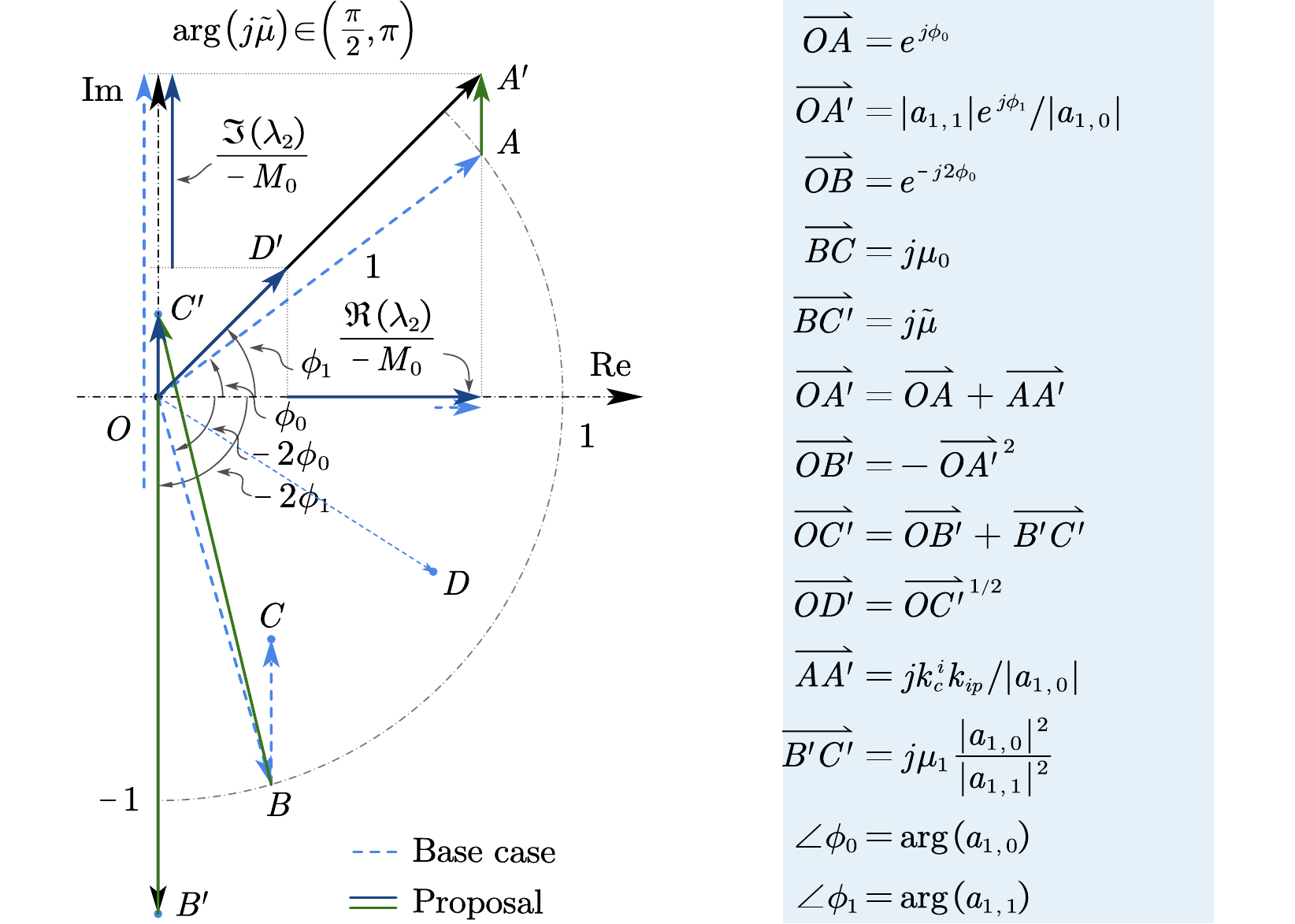}
            \captionsetup{singlelinecheck = false, format= plain, justification=justified, font=footnotesize, labelsep=period}
            \caption{Proposed complex intermediate parameter $\tilde{\mu}$ and its impact on varying the dominate pole $\lambda_2$ of the voltage loop. Base case quantities have subscript `0', and proposed quantities have subscript `1'. Points $A,B,C,D$ of the base case move to $A',B',C',D'$ due to the introduction of the imaginary part of the complex current feeding gain $k_c^i$, respectively.}
            %\vspace{-0.5cm}
            \label{fig:3.1_V_voltage_group_proposal}
        \end{figure}
        The equivalent vector $j\tilde{\mu}$ has a desired vectorial variation, improving the damping of $\lambda_2$ significantly.
        It can be achieved by introducing complex coefficients by letting $\Im({\tilde{k}_c})>0$ or $\Im({\tilde{k}_{vi}})<0$.
        % It agrees with the conclusion that complex coefficients show robustness improvements to controllers \cite{doria2016design}.
        In this design, the complex ${\tilde{k}_c} = k_c^r + j k_c^i \in \mathbb{C} $ is utilized, as it has minimum impact on the original control philosophy and is easy to modify. The implementation is setting the original coefficients with 
            \begin{align}
                &\beta_k \coloneqq k_c^r, &\beta_v \coloneqq -{j} k_c^i,
            \end{align}
        Fig.~\ref{fig:3-VoltCL_Ctrl_Proposed} depicts the implementation of the complex current feeding gain $\tilde{k}_c$ in the VC loop of Fig.~\ref{fig:0-SysDiagram}, the original diagram with $B_f$ and $k_{vp}$ being 0 to align with the assumption of (\ref{eq:7}). 
        Fig.~\ref{fig:3-VoltCL_Proposed} shows the SSM with $\tilde{k}_c$ variation. To include coupling terms, $a_1$ and $\mu$ in (\ref{eq:12_},\ref{eq:M, mu}) are updated to
        \begin{align}
            a_{1,1}&=k_{c}^{r}k_{ip}+L_gk_{ip}k_{vi}+j\left( X_g+k_{c}^{i}k_{ip} \right) , \\
            \label{eq:mu_new}
            \mu_1 &=\frac{4k_{ip}}{|a_1|^2}\left[ X_g\left( k_{c}^{r}-k_{vi}L_s \right) +\left( k_{c}^{r}+k_{vi}L_g \right) k_{c}^{i}k_{ip} \right] .
        \end{align}

        \begin{figure}[!]
            \centering
                {\includegraphics[width=1\linewidth]{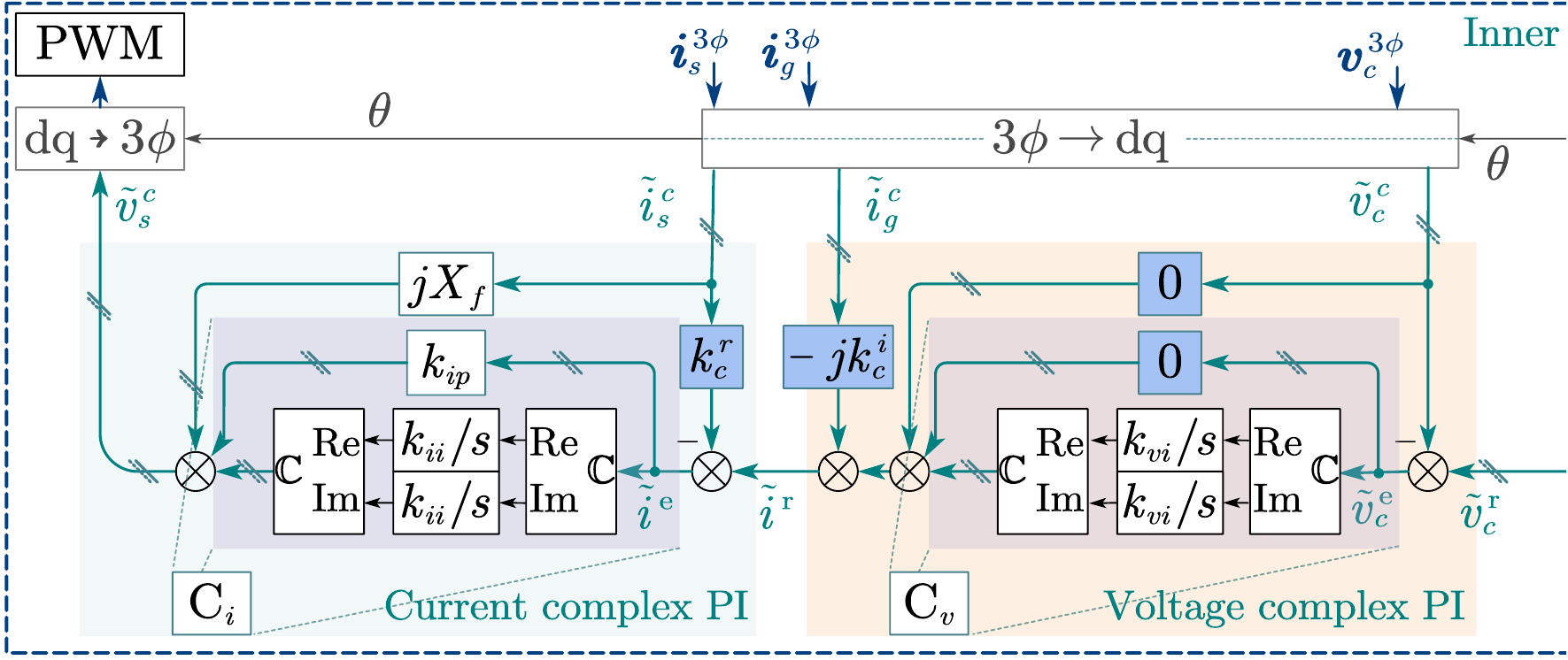}}
                \captionsetup{singlelinecheck = false, format= plain, justification=justified, font=footnotesize, labelsep=period}
                \caption{Revised control implementation of the proposed complex current feeding gain in the voltage-current loop.}
                \label{fig:3-VoltCL_Ctrl_Proposed}
        \end{figure}
        \begin{figure}[!]
            \centering
                {\includegraphics[width=1\linewidth]{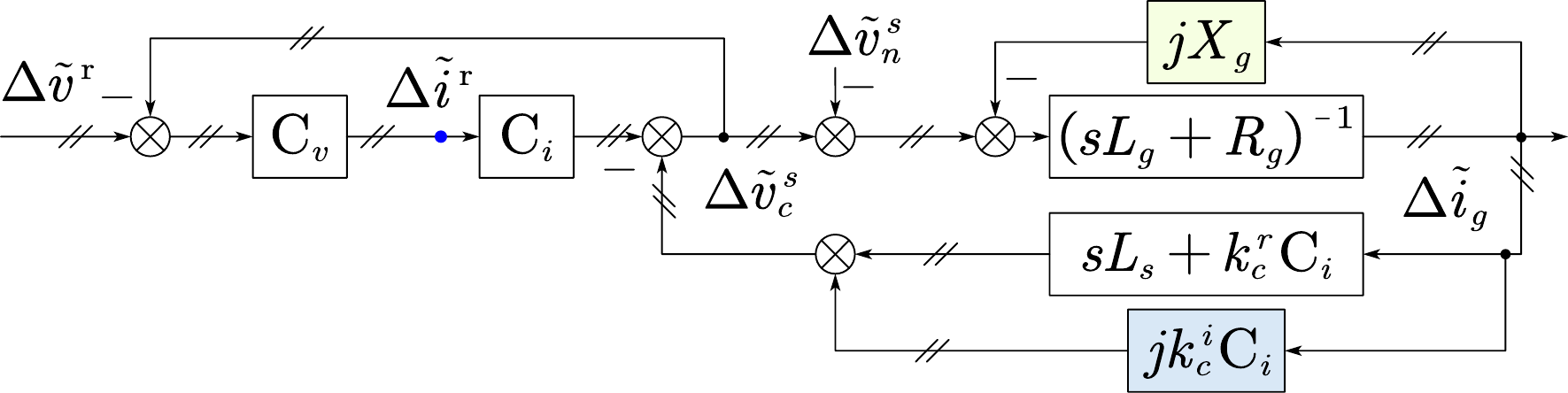}}
                \captionsetup{singlelinecheck = false, format= plain, justification=justified, font=footnotesize, labelsep=period}
                \caption{Modified complex voltage loop of VSG with complex current feeding gain $\tilde{k}_c$ and grid dynamics. The proposed branch is shadowed in blue.}
                %\vspace{-0.7cm}
                \label{fig:3-VoltCL_Proposed}
        \end{figure}
        The control tuning process can be complex and involve much trial and error, due to the dependency of intermediate parameters of $\lambda_2$. Two methods are hereby proposed. 
        
        \subsubsection{Tuning by Optimization}\label{sec:3-1-2}
        An optimization problem is formulated to 
        minimize the integral squared error (ISE) of the control magnitude response to the reference \cite{skogestad2005multivariable} defined as $T_{\mathrm{ref}}(s)=(0.005s+1)^{-1}$, targeting 20~ms rise time.
        \begin{mini}
            {\tilde{k}_c}{\mathrm{ISE}=\small{\sqrt{\int_{0.01}^{0.06}{\left( | {\tilde{G}_{v,\mathrm{cl}}(\tilde{k}_c,t)} |-| T_{\mathrm{ref}}(t) | \right)^{2} dt}}},}{}{}
            \addConstraint{\tilde{k}_c \in \mathbf{A},}
        \end{mini}
        where $\mathbf{A}$ is the possible range of $\tilde{k}_c$. The integration interval focuses on the dominant dynamic period. Fig.~\ref{fig:k_c_contour_ISE_min}(a) depicts an overlapped contour of the natural frequency $\omega_{n}$ and damping ratio $\zeta$ of $\lambda_2$ by solving (\ref{eq:v_poles}). The grey shaded area $\mathbf{A}$ sets the boundary of $\tilde{k}_c$, where $\omega_{n}>100$~rad/s and $\zeta>0.7$.
        \begin{figure}[!]
            \centering
            {\subfloat[]
                {\includegraphics[width=0.49\linewidth]{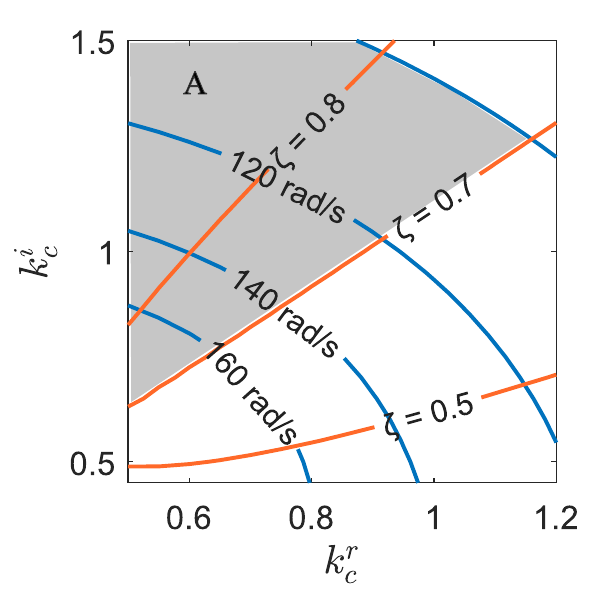}}}
            {\subfloat[]
                {\includegraphics[width=0.49\linewidth]{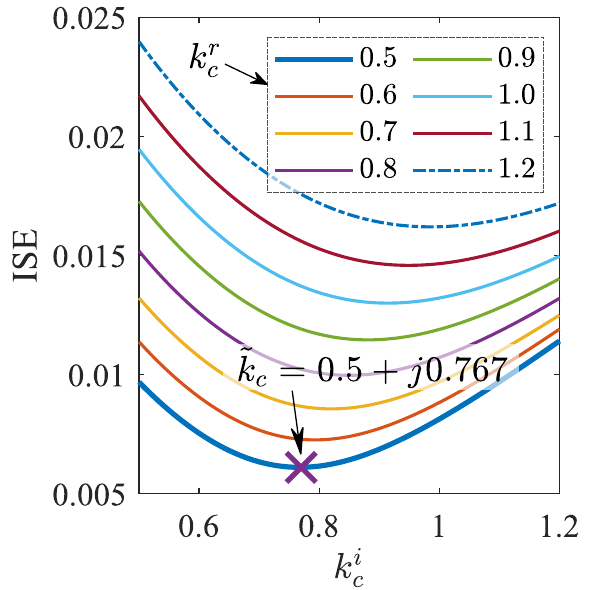}}}
            \captionsetup{singlelinecheck = false, format= plain, justification=justified, font=footnotesize, labelsep=period}
            \caption{(a) The overlapped contours of the natural frequency and damping ratio of the dominant pole $\lambda_2$ by varying $\tilde{k}_c$ on a complex plane. The shaded area marked $A$ is a potential selection area. (b) The ISE variation of the potential loop response to the reference with $\tilde{k}_c$ variation.
            }
            %\vspace{-6mm}
                \label{fig:k_c_contour_ISE_min}            
        \end{figure}
        {The minimized ISE solution is $\tilde{k}_c = 0.5+j0.767$ as marked in Fig.~\ref{fig:k_c_contour_ISE_min}(b) with $\tilde{k}_c \in \mathbf{A}$.} The resultant $\lambda_2=176 \angle 218.6^{\circ}~\mathrm{s}^{-1}$ corresponds to a 15~ms rise time and a 2.74\% overshoot. 
        
        \subsubsection{Direct Pole-placement}\label{sec:3-1-3}
        This approach avoids tedious optimization and provides a straightforward procedure.
        It directly places the dominant pole with proper damping, i.e., $\zeta = 0.707$,
        by using the special geometrical feature of (\ref{eq:v_poles, alt}).
        {Let $\phi = 45^{\circ}$, then $\mathrm{arg}(e^{-j2\phi}) = -90^{\circ}$ and $\mathrm{arg}(j\mu) = 90^{\circ}$. Therefore, $|\mu|>1$ ensures the argument of the angle dominant part to be $45^{\circ}$ aligning with $(e^{j\phi})$, as shown in Fig.~\ref{fig:3.1_V_voltage_group_proposal}.} In this way, two poles share $\zeta = 0.707$, placing at
        \begin{align}
            \label{eq:pole-placement}
            \lambda _{1,2}=-\frac{|a_1|}{2a_2}\left[ 1\pm \left( 1-4\frac{|a_0|a_2}{|a_1|^2} \right) ^{\frac{1}{2}} \right] e^{j\frac{\pi}{4}},
        \end{align}
        when
        \begin{align}
            \label{eq:kc_placement}
            \tilde{k}_{c}=k_{c}^{r}\left( 1+j \right) +j\left( L_gk_{vi}-X_g/k_{ip} \right).
        \end{align}
        By keeping $k_{c}^{r}=\beta_k=1$, the varied $\tilde{k}_{c}$ is $1+j1.1356$, and the pole $\lambda_2 = 110 \angle 225^{\circ}~\mathrm{s}^{-1}$. The time-domain response has a rise time of 19.7 ms and an overshoot of 4.63\%. The performance is within the specification.
        
        \begin{figure}[!]
            \centering
            \includegraphics[width=1\linewidth]{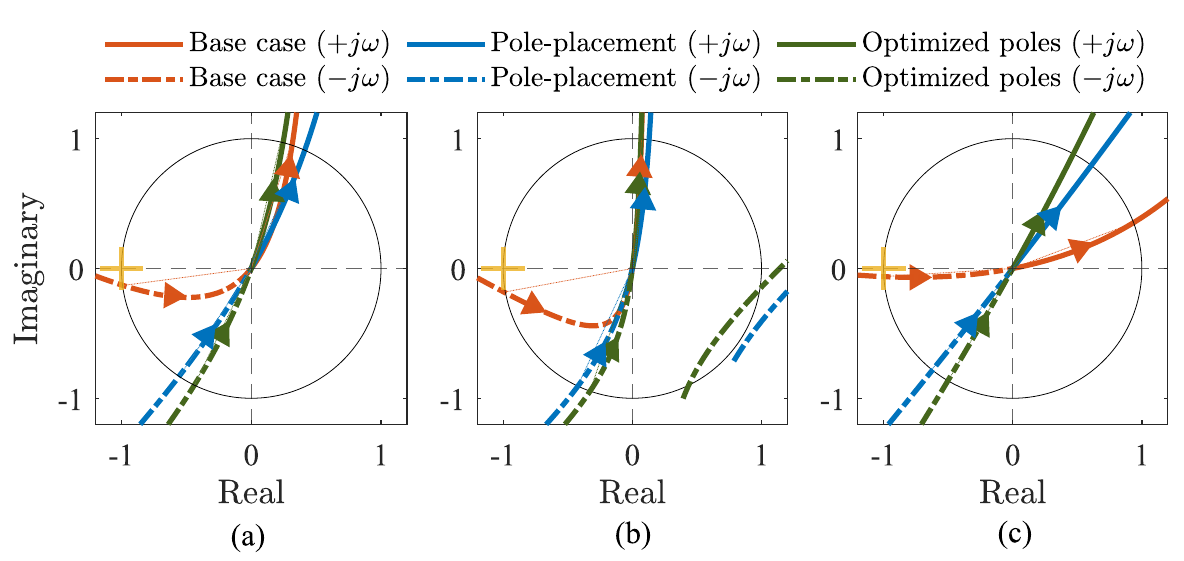}
            \captionsetup{singlelinecheck = false, format= plain, justification=justified, font=footnotesize, labelsep=period}
            \caption{
            The Nyquist plot of the complex voltage minor-loop coupled with various grid impedance:
            (a) $X_g = 0.30$, 
                    the positive frequency phase margins are $\left\{ \phi _{m,+j\omega} \right\} =\left\{ 108.4^{\circ},114.4^{\circ},104.2^{\circ} \right\} $,
                    the negative frequency phase margin are $\left\{ \phi _{m,-j\omega} \right\} =\left\{ 7.6^{\circ},57.2^{\circ},64.0^{\circ} \right\} $;
            (b) $X_g = 1.50$,
                    $\left\{ \phi _{m,+j\omega} \right\} =\left\{ 94.0^{\circ},97.4^{\circ},93.8^{\circ} \right\} $,
                    $\left\{ \phi _{m,-j\omega} \right\} =\left\{ 11.0^{\circ},65.0^{\circ},71.0^{\circ} \right\} $;
            (c) $X_g = 0.04$,
                    $\left\{ \phi _{m,+j\omega} \right\} =\left\{ 159.4^{\circ}, 127.4^{\circ}, 114.7^{\circ} \right\} $,
                    $\left\{ \phi _{m,-j\omega} \right\} =\left\{ 3.6^{\circ}, 51.6^{\circ}, 60.0^{\circ} \right\} $;
            where sets have the order for the base case, the pole placement setup, and the optimized pole setup.
            }
            %\vspace{-0.7cm}
            \label{fig:3._Nyquist_Vc}
        \end{figure}
        
        \subsubsection{Robustness Validation}\label{sec:3-1-4}
        To determine the effectiveness and robustness of the design proposal, a few comparisons are provided. 
        Fig.~\ref{fig:3._Nyquist_Vc} compares the phase margin of the complex voltage minor-loop gain in Nyquist plots
        \footnote{The phase margin in the negative frequency region ($-j\omega$) is the angle from $180^{\circ}$ to the intersection of the trace to the unit circle in the clockwise direction, whereas in the positive frequency region ($j\omega$) it is the angle from $-180^{\circ}$ to the trace intersection in the counter-clockwise direction.} 
        \cite{doria2018complex}
        with three grid conditions $X_g =\{0.3, 0.04, 1.5\}$ via the inverse of the open-loop CTF (\ref{eq:11a}). 
        The Nyquist plot suggests that the negative frequency phase margins are smaller than the positive phase margins in general, {hence, determining the voltage loop dynamics \cite{bishop2011modern}}.
        Furthermore, the $\tilde{k}_c$ embedded cases exhibit superior phase margins than the real gain base case. In three cases, $\tilde{k}_c$ guarantees a minimum $50^{\circ}$ phase margin compared to the margin $\le 10^{\circ}$ of the base case, which further drops to $3.6^{\circ}$ if coupled with a strong grid. The optimized pole choice has marginally larger phase gains than the pole-placement choices.  

        \par Fig.~\ref{fig:3.2_rlocus_Varyk_vi} shows the root loci of the complex voltage loop to the variation of voltage integral gain $k_{vi}$.
        Fig.~\ref{fig:3.2_rlocus_Varyk_vi}(a) depicts scenarios with the real feeding gain, i.e. $k_c \in \Re$. A positive $\beta_v$ can improve the damping. However, a too-high $k_{vi}$, as the base case, leads $\zeta \rightarrow 0$. Further increasing $k_{vi}$ pushes the pole to the right-half plane. A fully compensated current feedback gain with $\beta_v=1$ or equivalently $k_c=0$ has right-half plane poles at all times.
        Fig.~\ref{fig:3.2_rlocus_Varyk_vi}(b) shows complex feeding gains with a positive imaginary part, i.e. $k_c^i>0$, exhibits superior robustness. The pole stays in the left-half plane against a vast variation of $k_{vi}$. Both the speed and damping of the pole are extensively improved compared to real $k_c$ cases. In contrast, a negative $k_c^i$ possibly causes the system to be unstable. Fig.~\ref{fig:3.2_rlocus_Varyk_vi}(a) also overlays the root loci of the comprehensive SSM upon the complex voltage loop loci of the base case. Two trajectories closely match each other.

        \begin{figure}[!]
            \centering
                {\includegraphics[width=1\linewidth]{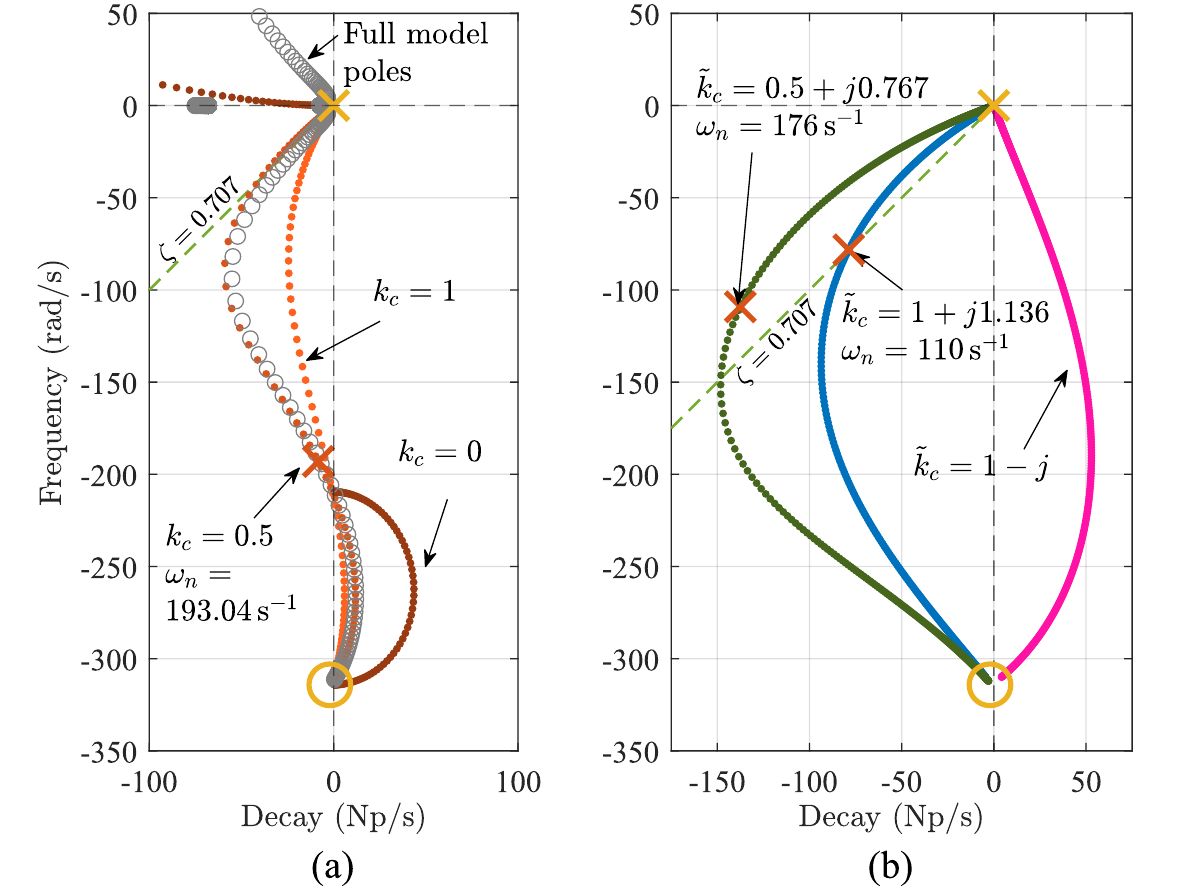}}
                \captionsetup{singlelinecheck = false, format= plain, justification=justified, font=footnotesize, labelsep=period}
                \caption{
                Root loci comparison of the complex voltage loop with different current feeding gains when varying $k_{vi}\in[8, 80000]$:
                (a) the base case $k_c = 0.5$ with overlapped full model poles, and scenarios with $k_c=\{0,1\}$;
                (b) the complex feeding gain $\tilde{k}_c$ with the optimized value $0.5+j0.767$, the pole-placement value $1+j1.136$ and unstable scenario $1-j$.
                }
                \label{fig:3.2_rlocus_Varyk_vi}
        \end{figure}
        \begin{figure}[!]
            \centering
                {\includegraphics[width=1\linewidth]{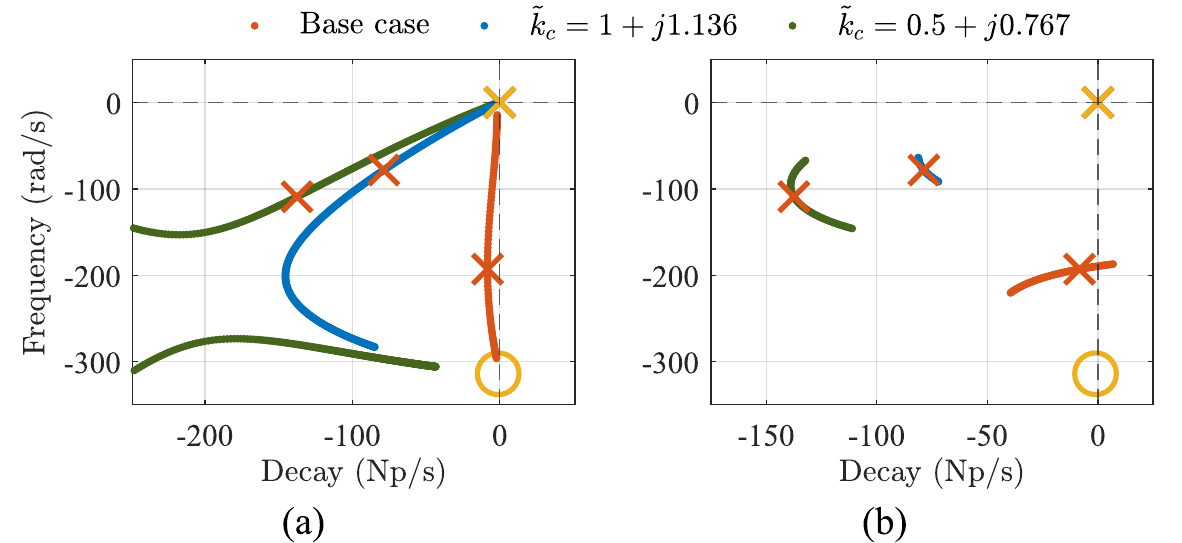}}
               \captionsetup{singlelinecheck = false, format= plain, justification=justified, font=footnotesize, labelsep=period}
                \caption{
            Root loci comparison of the complex voltage loop with different current feeding gains $\tilde{k}_c\in\{0.5, 1+j1.136, 0.5+j0.767\}$:
            (a) Varying grid impedance $X_g \in [10^{-3}, 1.5]$ p.u.,
            (b) Varying filter impedance $X_s \in [0.01, 0.30]$ p.u..
            }
            %\vspace{-0.5cm}
            \label{fig:3.3_rlocus_Xg_delX}
        \end{figure}
        
        Fig.~\ref{fig:3.3_rlocus_Xg_delX} depicts the root loci comparison of the system facing grid or filter uncertainties, where control parameters remain the same. Briefly, both scenarios demonstrate the complex current gain keeps the loop stable.
        Fig.~\ref{fig:3.3_rlocus_Xg_delX}(a) shows $\tilde{k}_c$ maintains proper damping against a wide range variation of grid impedance, e.g. $\zeta > 0.56, \forall X_g<0.86$ with $\tilde{k}_c=1+j1.1356$. The small grid impedance sacrifices the control speed though, e.g. $|\lambda_2|=19.8$~s$^{-1}$ if $X_g = 0.04$.
        Fig.~\ref{fig:3.3_rlocus_Xg_delX}(b) shows the stiffness of the dominant pole $\lambda_2$ against the filter uncertainty. With the $\tilde{k}_c$ application, the usual ${k}_c^r=1$ setting has the least pole variation, whereas a real ${k}_c$ can lead to instability.

        These examinations conclude the proposed complex current feeding gain solution maintains the system's stability from a wide range of uncertainty with assured damping. The proposed direct pole placement method shows its tuning simplicity. $\tilde{k}_{c}=1+j1.1356$ is to be used unless otherwise specified.
        %\vspace{-4mm}

    \subsection{Voltage-Caused Power Dynamics Compensation} \label{sec:3-2} %\vspace{-2mm}
        \par The previously proposed approach tunes the voltage loop to being capable of holding the output magnitude rigorously. Meanwhile, in the voltage-loop-dominant time scale, the $dq\leftrightarrow 3 \phi$ induced coupling is desired to be suppressed. 
        The compensation strategy is to reverse the identified angle rotating term $G_{\theta ,v}$ via
        \begin{align}
            \label{eq:3.3.1}
            \mathrm{C}_{v,\theta} &= -\frac{j\Delta \theta}{\Delta \tilde{v}^{\mathrm{e}}}=\mathrm{C}_v\mathrm{C}_iG_{\theta ,v}^{-1}\IEEEnonumber
            \\
            &=\left[ \tilde{v}_{c}^{0}\left( \mathrm{C}_v\mathrm{C}_i \right) ^{-1}+\tilde{v}_{c}^{0}+\tilde{k}_c\tilde{i}_{g}^{0}{\mathrm{C}_v}^{-1} \right] ^{-1}\IEEEnonumber
            \\
            &\approx \frac{1}{1+ ( k_{ip}k_{vi})^{-1}s}.
        \end{align}
        The approximation holds with the terminal voltage around 1.0 p.u. and negligible steady-state currents \cite{harnefors2020universal}. In addition, $\Delta \tilde{v}_c^{\mathrm{e}} = \Delta {v}_{\mathrm{d}}^{\mathrm{e}}$ holds for ${v}_{c,\mathrm{q}}^0=0$. {The correctional power angle is, therefore, inversely proportional to the voltage magnitude error, which correlates to (\ref{eq:12}) when $\Delta P = 0$.} {The error present drives the compensator to primarily apply a reverse synthetic electrical torque to the inverter's frame.} The varied voltage CTF is 
        \begin{align}
            \label{eq:11c}
            \frac{\Delta \tilde{v}_c}{\Delta \tilde{v}^{\mathrm{r}}-\Delta \tilde{v}_c}=\frac{\mathrm{C}_i\mathrm{C}_v - j{\mathrm{C}_{v,\theta}}  }{L_P+1}.
        \end{align}
        \begin{figure}[!]
            \centering
                {\includegraphics[width=1\linewidth]{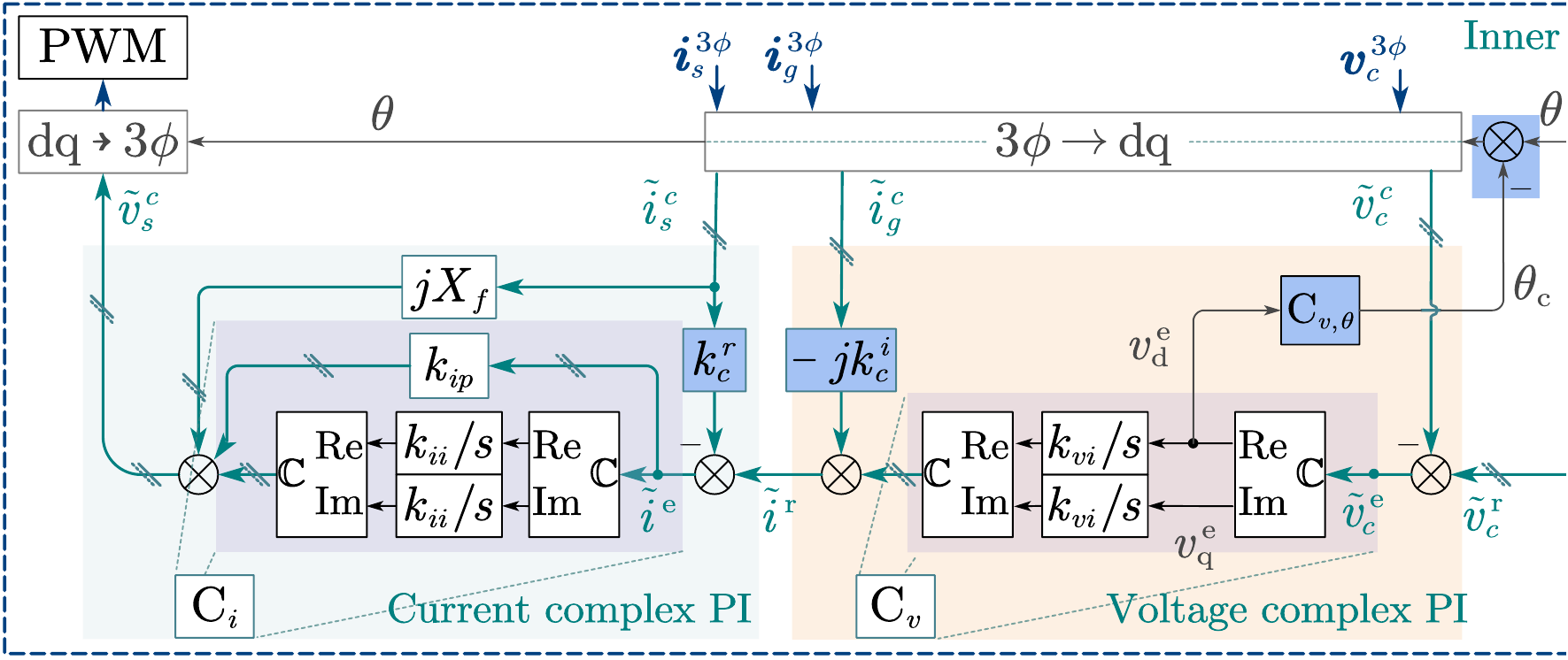}}
                \captionsetup{singlelinecheck = false, format= plain, justification=justified, font=footnotesize, labelsep=period}
                \caption{The modified control diagram with complex current feeding gain and voltage-angle compensator. Changes are shaded in blue.}
                \label{fig:0-SysDiagram_compensation}
                %\vspace{-.6cm}
        \end{figure}
        \begin{figure}[!]
            \centering
                {\includegraphics[width=1\linewidth]{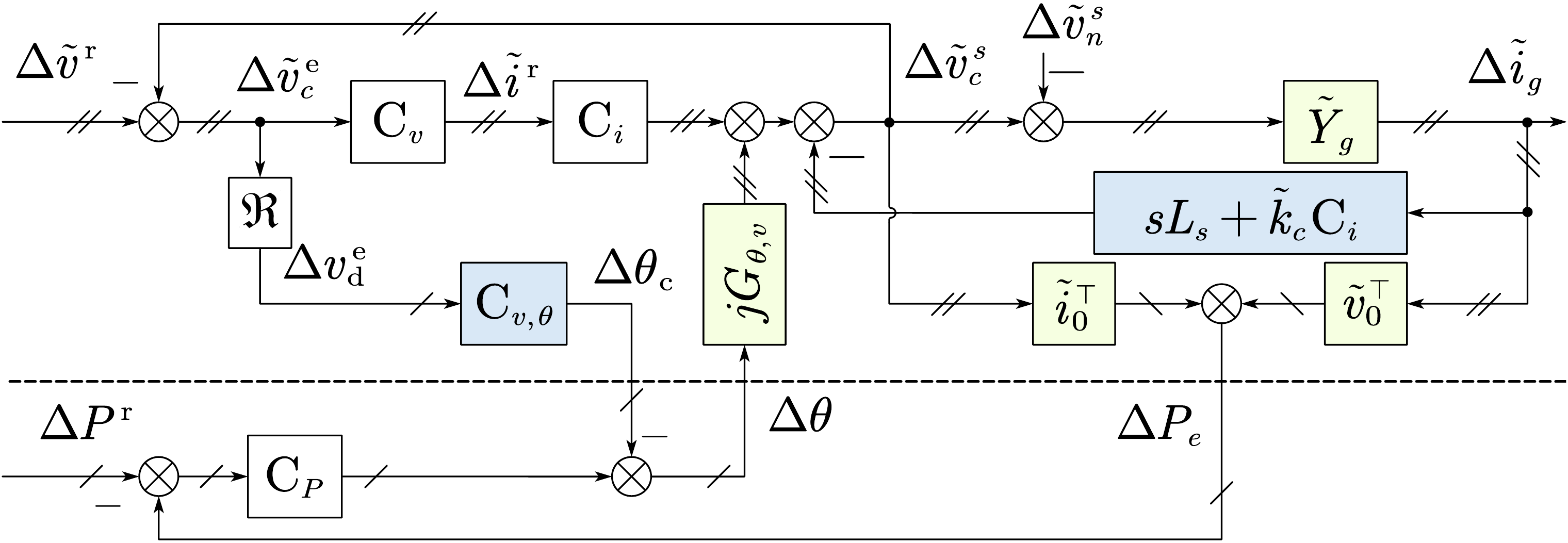}}
                \captionsetup{singlelinecheck = false, format= plain, justification=justified, font=footnotesize, labelsep=period}
                \caption{Modified small-signal model of the VSG with amended complex grid feeding gain and the voltage-angle compensator.}
                \label{fig:3.1-PowrCL-Proposed}
                %\vspace{-.6cm}
        \end{figure}
        \begin{figure}[!]
            \centering
            {\includegraphics[width=0.75\linewidth]{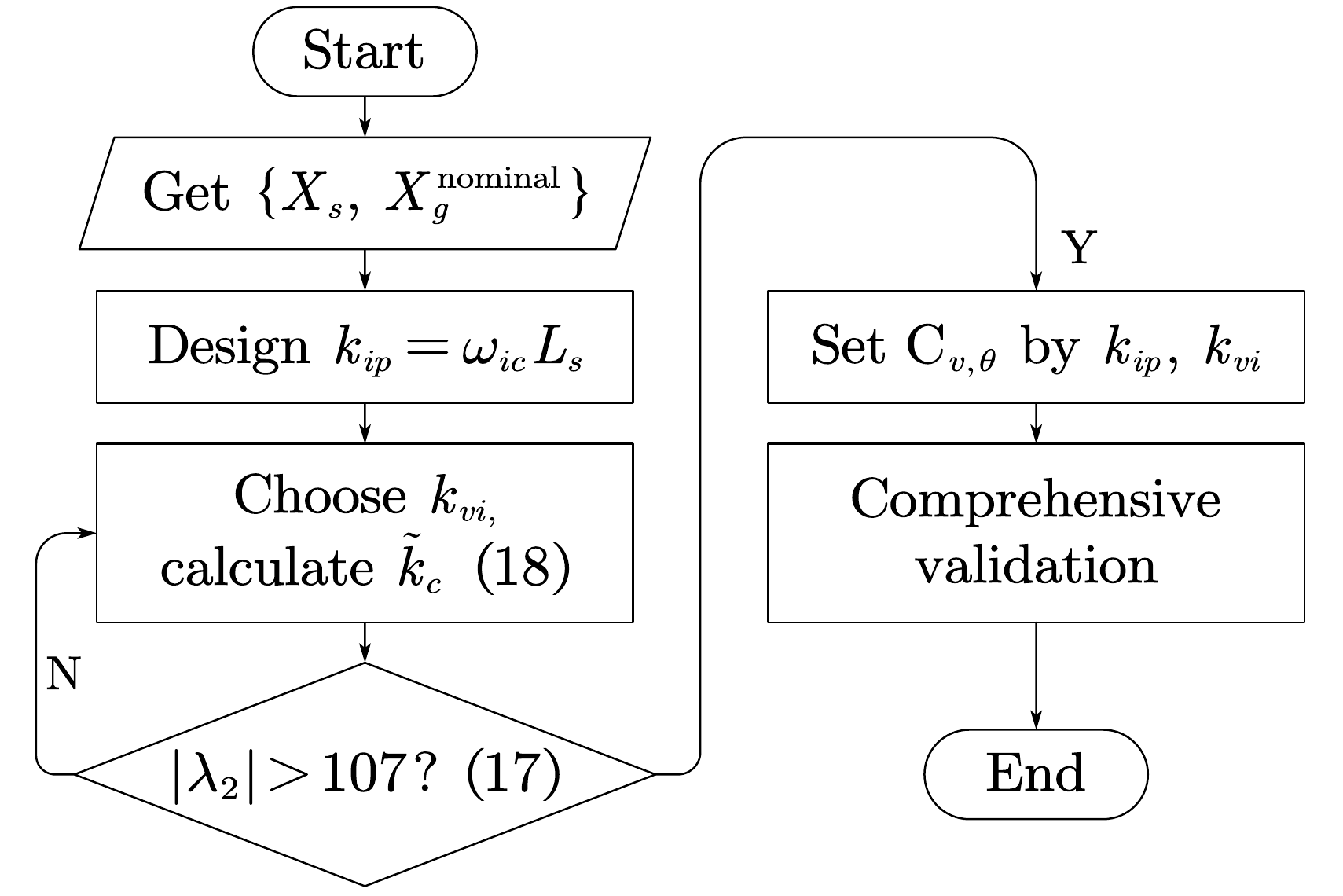}}
            \captionsetup{singlelinecheck = false, format= plain, justification=justified, font=footnotesize, labelsep=period}
            \caption{The control parameter design flow chart.}
            \label{fig:ParamDesingFlowChart}
            %\vspace{-.5cm}
        \end{figure}
    Fig.~\ref{fig:0-SysDiagram_compensation} depicts the varied control loop implementation.
        Fig.~\ref{fig:3.1-PowrCL-Proposed} depicts the varied SSM block diagram with $\mathrm{C}_{v,\theta}$ term.
        The overall control design procedure is summarized as follows, with the associated flow chart being Fig.~\ref{fig:ParamDesingFlowChart}:
        \begin{enumerate}
            \item  Get system information with the filter reactance $X_s$ and the lowest nominal grid reactance $X_g$ in p.u., including step-up transformers,
            \item  Choose desired current open-loop gain $\omega_{ic}$,
            \item  Choose a voltage integral gain $k_{vi}$, and
             calculate the complex gain $\tilde{k}_c$ by (\ref{eq:kc_placement}) to set $\zeta = 0.707$,
            \item  Iterate $k_{vi}$ until the pole natural frequency $|\lambda_2|>107$~rad/s by (\ref{eq:pole-placement}), keeping 20~ms rise time by (\ref{eq:v_step_approx}),
            \item  Set compensator $\mathrm{C}_{v,\theta}$ by desired $k_{vi}$ and $k_{ip}$,
            \item  Test the designed control as per Fig.~\ref{fig:0-SysDiagram_compensation}. {Further, validate the full VSG scheme comprehensively.}
        \end{enumerate}

    This two-fold design has its unique advantages. The tuning needs minimum inputs with guaranteed performance and stability margin against vast uncertainties. The complexity and computational burden are much reduced. Furthermore, the implementation has minimized variation to the generic VSG control scheme. %\vspace{-0.5cm}

        \subsection{The Collective Impact of Complex Current Feeding Gain and Voltage-Angle Compensator on vSSCI Resonance} \label{sec:3-3}
       
        \par 
        Fig.~\ref{fig:Freq_Vds2Vdr+Pe2Vdr}(a) compares the closed-loop frequency response (CLFR) of PoC $d$-axis voltage to its reference change, within the original design, the complex current feeding gain $\tilde{k}_c$, and with added voltage-angle compensator. The base case forms a vSSCI resonance peak at 30.37~Hz, of which the oscillation mode is identified in Fig.~\ref{fig:2.0-VoltResponseComp}. The resonance peak is rectified by embedding $\tilde{k}_c$, and the closed-loop gain (CLG) at -3~dB is 18.0~Hz. The reshaped CLFR lies in the envelope formed by the positive and negative frequency responses of $\tilde{G}_{v,\mathrm{cl}}$, where the grid impedance zero causes the notch.
        The addition of $\mathrm{C}_{v.\theta}$ extends the CLG to 21.6 Hz.
        \begin{figure}[!]
            \centering
            \includegraphics[width=1\linewidth]{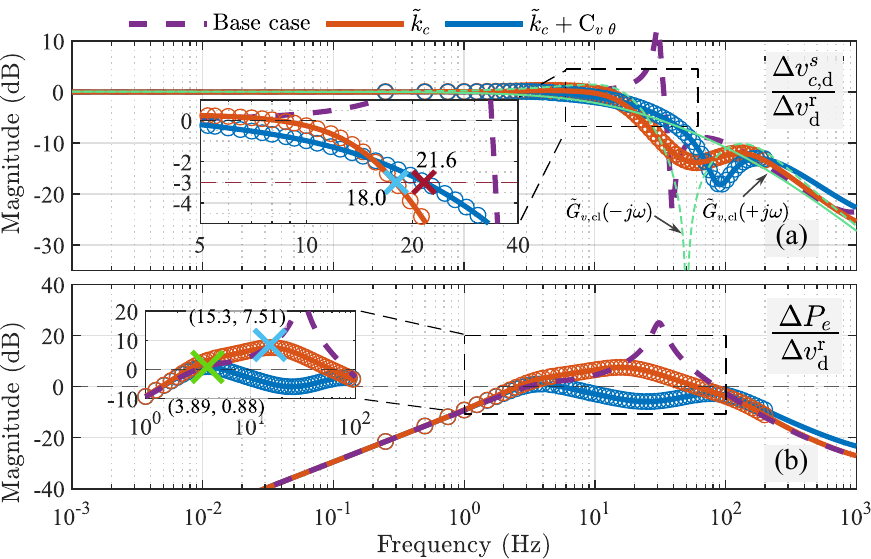}
            \captionsetup{singlelinecheck = false, format= plain, justification=justified, font=footnotesize, labelsep=period}
            \caption{Comparison of CLFRs to the voltage reference ${\Delta v_{\mathrm{d}}^\mathrm{r}}$ perturbation of a VSG being configured as the base case, with complex gain $\tilde{k}_c$, and added with voltage-angle compensator $\mathrm{C}_{v,\theta}$:          
            (a) PoC $d$-axis voltage ${{\Delta v_{c,\mathrm{d}}^{s}}}$ and (b) active power ${\Delta P_e}$. Circles depict the identified frequency responses. The positive and negative frequency responses of the complex loop $\tilde{G}_{v,\mathrm{cl}}$ are overlapped.}
            \label{fig:Freq_Vds2Vdr+Pe2Vdr}
            %\vspace{-0.5cm}
        \end{figure}
        
        Fig.~\ref{fig:Freq_Vds2Vdr+Pe2Vdr}(b) compares the CLFR of real power to voltage reference change, with the same configuration. The original design has the same resonance mode. The peak is compensated by $\tilde{k}_c$, but still, a positive dB region of coupling is found ($f\in[2.6, 60]$ Hz). The maximum amplification is 7.51~dB (2.37) at 15.3~Hz.
        This region is suppressed by adding $\mathrm{C}_{v,\theta}$. The maximum amplification is 0.88~dB (1.11) at 3.89~Hz. $\mathrm{C}_{v,\theta}$ works as a notch filter. 
        
        The above theoretical CLFRs have been verified by the response identification in simulations. A summed multi-tone sinusoidal perturbation is injected into ${\Delta v_{\mathrm{d}}^\mathrm{r}}$. The recorded $v_{c,\mathrm{d}}^s$ and $P_e$ are processed by discrete Fourier transform (DFT) and compared to the DFTed perturbation, yielding the identified frequency responses depicted by circles.

        In summary, the complex current feeding gain rectifies the {vSSCI} resonance peak. The added voltage-angle compensator extends the voltage CLG and suppresses the voltage-induced power coupling by nearly 0 dB. 
        %\vspace{-.2cm}

\section{Performance Evaluation} \label{sec:4}
This section validates the terminal dynamic performance of the design in time-domain simulation and experiments with the enhancement provided by the complex current feeding gain and voltage-angle compensator independently and jointly. Furthermore, the robustness of the proposal against strong grid, phase jump, and frequency drop is also validated. The simulation gives a performance comparison with the state-of-the-art solution, {active susceptance (AS) \cite{zhao2022robust}}. It also tests the design efficacy in the SG-dominant grid on the IEEE 9-bus bench system. 
% %\vspace{-.5cm}

{\subsection{Simulation Verification}}
    To verify the efficacy of the proposal, a 0.1 p.u. step on set points is applied on various VSG configurations, respectively. Terminal responses perturbed by (a) $\Delta \tilde{v}^{\mathrm{r}}$ and (b) $\Delta P^{\mathrm{r}}$ are depicted as Fig.~\ref{sim_Verification}.
    VSGs are configured with (i) lowered gain $k_{vi}=200$ for $\zeta=0.643$, as per Fig.~\ref{fig:3.2_rlocus_Varyk_vi}(a), (ii) the lowered $k_{vi}$ and added compensator $\mathrm{C}_{v,\theta}$, (iii) the proposed complex gain $\tilde{k}_{c}$ on the base case, (iv) the combined variation of $\tilde{k}_{c} + \mathrm{C}_{v,\theta}$ on the base case, and (v) the AS voltage-current control scheme \cite{zhao2022robust}, sequentially. To compare fairly, all controllers keep a 20~ms voltage rise time. Power loops and circuits remain the same as the base case.

    Fig.~\ref{sim_Verification}(a) depicts terminal responses, $ v_{c,\mathrm{d}}$, $I_g$, $P_e$, $Q_e$, and $\Delta \theta$, during a voltage transition. Cases (iv) and (v) have the least voltage overshoot. $\tilde{k}_{c}$ shows its significant damping improvement and limited overshoot as per design.
    $\mathrm{C}_{v,\theta}$ shows its current and power peak suppression capability by comparing case (ii) to (i) and case (iv) to (iii). The proposal, case (iv), limits the coupled active power trajectory to less than 0.1 p.u., the voltage step-change magnitude. However, $\mathrm{C}_{v,\theta}$ has direct angle manipulation and, therefore, leads to a transient angle during a voltage fluctuation. (iv) recovers from the sag within 20~ms as the voltage stabilizes. The AS scheme (v) has the optimized performance in this test.

     Fig.~\ref{sim_Verification}(b) depicts the same terminal responses after a real power transition. Cases (i-iv) show proper voltage controllability during the transition with 0.4\% of voltage sag, while case (v) shows a 3.4\% voltage spike due to the relatively higher $X_g$ than in \cite{zhao2022robust}.
     In summary, the design targets of terminal response enhancement have been initially checked in simulations. A high-speed and properly damped voltage loop is ensured by cooperatively tuning $k_{vi}$ and $\tilde{k}_{c}$. $\mathrm{C}_{v,\theta}$ shows its power suppression capability during voltage disturbances. A fast voltage control restrains the transient angle sag duration.
     % \vspace{-.5cm}
        \begin{figure}[!]
            \centering
            \includegraphics[width=1\linewidth]{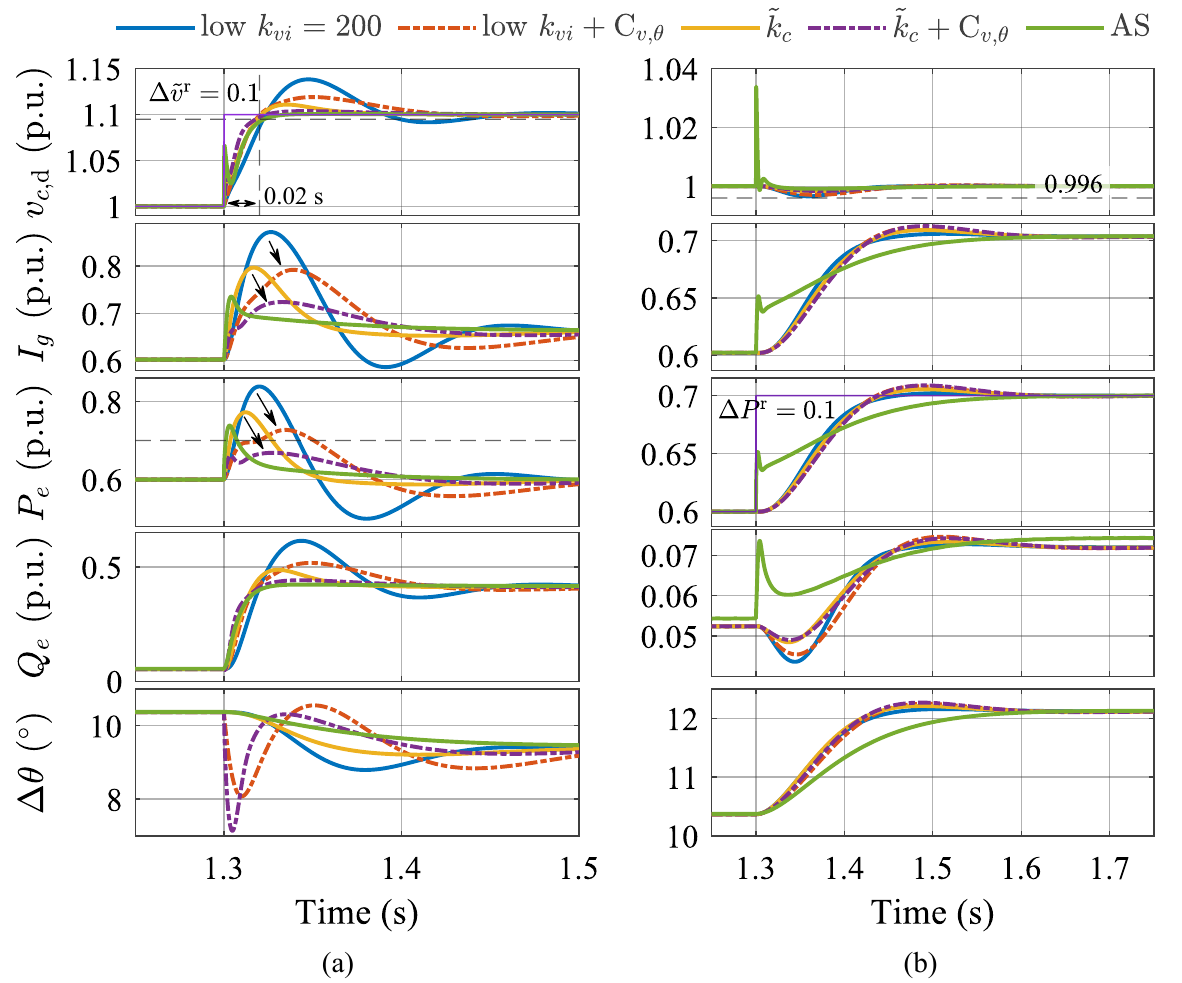}
            \captionsetup{singlelinecheck = false, format= plain, justification=justified, font=footnotesize, labelsep=period}
            \caption{Terminal responses comparison for a 0.1 p.u. reference change of the VSG SMIB base case with various configurations: (a) voltage $\Delta \tilde{v}^{\mathrm{r}}$, (b) active power $\Delta P^{\mathrm{r}}$, at 1.3 s. The base case configuration is varied where (i) reduced $k_{vi} = 200$, (ii) the low $k_{vi}$ added compensator $\mathrm{C}_{v,\theta}$, (iii) the base case with the proposed complex current feeding gain $\tilde{k}_{c}$, (iv) with additional $\mathrm{C}_{v,\theta}$, (v) active susceptance scheme \cite{zhao2022robust} ($k_{vi}= 420$) as the voltage-current loop.
            }
            \label{sim_Verification}
            % \vspace{-.5cm}
        \end{figure}

    {
    To validate the compatibility and efficacy of the proposed VSG control within an SG-dominant power grid, a set of tests are performed on the IEEE 9-bus system of which the bus naming and grid parameters are aligned with \cite{sauer2017power}. Three generators, SM1, SM2, and VSG, have their machine base of 500, 200, and 100~MVA, respectively. Synchronous machines SM1 and SM2 on bus 1 and 2 have the machine and exciter parameters labeled as BPS\_2 in \cite{vowles2015smallsignal}. The VSG on bus~3 coupled with $0.22$~p.u. Thévenin reactance is configured by (i) the proposed control, and (ii) with lowered $k_{vi}=200$. For brevity, the grid single-line diagram and machine parameters are omitted.
        \begin{figure}
            \centering
            % \colorbox{yellow}
            {\includegraphics[width=1\linewidth]{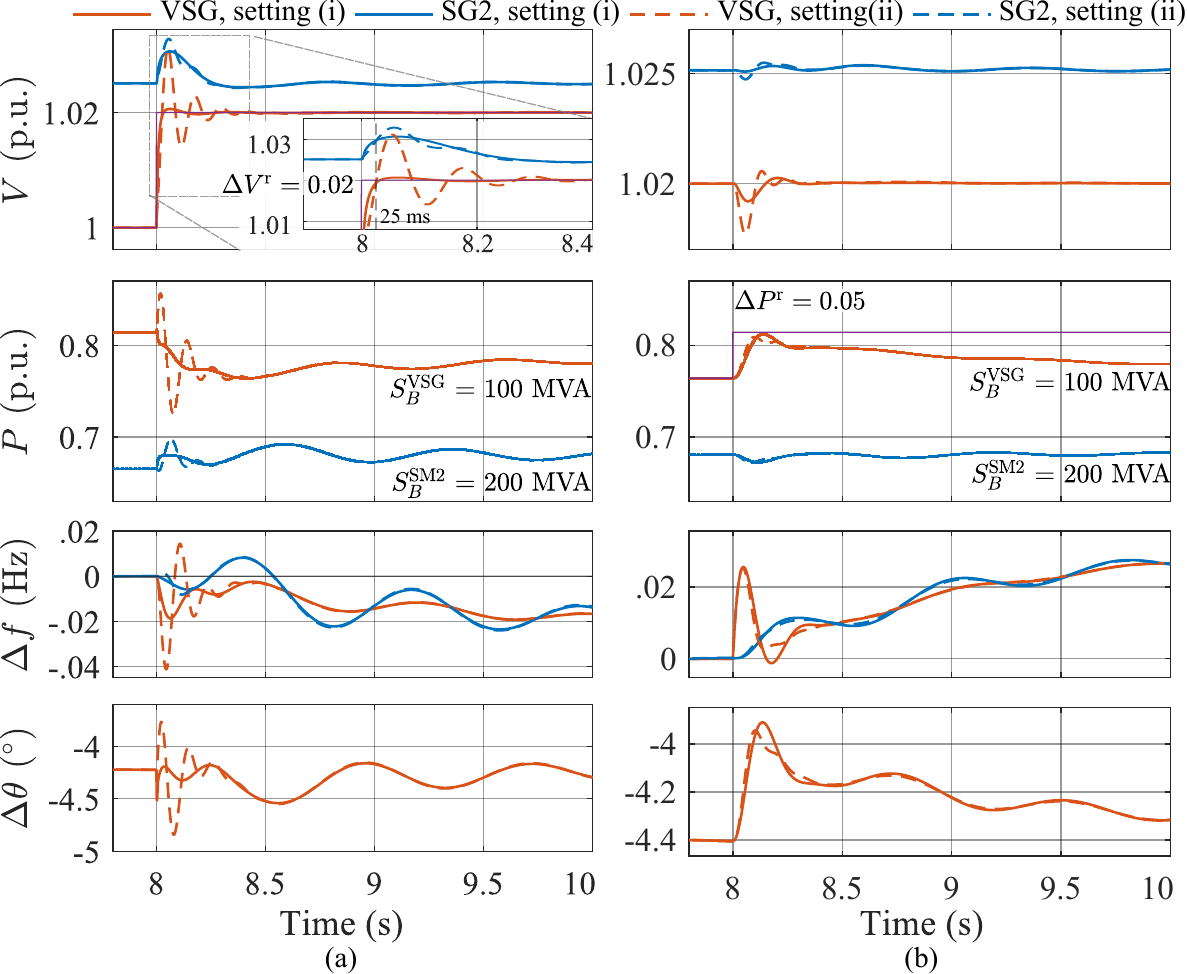}}
            \captionsetup{singlelinecheck = false, format= plain, justification=justified, font=footnotesize, labelsep=period}
            \caption{{IEEE 9-bus benchmark for (a) a 0.02~p.u. voltage reference step and (b) a 0.05~p.u. active power reference step at 8~s. The VSG is with the voltage loop settings (i) the proposed VSG control for damping ratio $\zeta=0.707$ and (ii) with lowered $k_{vi}=200$ for $\zeta=0.643$, at $X_g=0.3$.
            }}
            \label{fig:IEEE-9Bus-Test}
        \end{figure}

    Fig.~\ref{fig:IEEE-9Bus-Test} compares the SM1 and VSG terminal responses when perturbing (a) a 0.02 p.u. voltage reference change and (b) 0.05 p.u. power reference change on VSG. In both cases, the fast decaying voltage dynamics overlap the slower power dynamics. The proposed control keeps its robustness in speed and damping even against 36\% of SCR change. In contrast, the conventional VSG control deteriorates its voltage loop damping noticeably. The frequency deviation compensates for the power reference deviation as SG prime mover governors are not modeled. Overall, it shows the capability of the designed control to regulate its terminal voltage and decouple its impact on power-frequency dynamics. 
    }

{\subsection{Experimental Verification}}

    To validate the efficacy of the proposed terminal dynamic performance enhancement strategies, a scale-down experimental composed of an Imperix inverter, a constant voltage DC supply, and a Regatron AC grid emulator, as shown in Fig.~\ref{fig:ExpSetup}, is used. The varied circuit and control parameters from TABLE~\ref{tab:Circuit_and_Controller_Parameters_of_the_VSG} are listed in TABLE~\ref{tab:exp_setup}.
        \begin{figure}[!]
            \centering
            \includegraphics[width=1\linewidth]{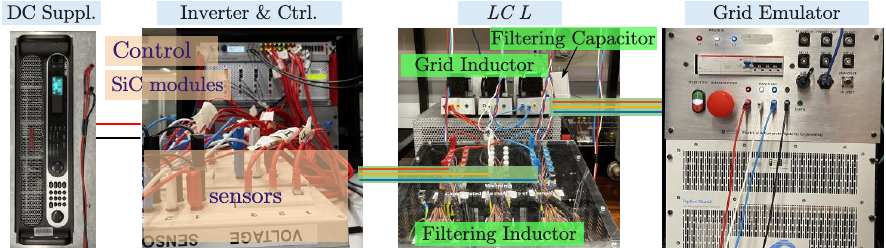}
            \captionsetup{singlelinecheck = false, format= plain, justification=justified, font=footnotesize, labelsep=period}
            \caption{The scaled-down experiment setup.
            }
            \label{fig:ExpSetup}
            %\vspace{-.35cm}
        \end{figure}

    \begin{table}[]
    	\centering
    	\captionsetup{singlelinecheck = false, format= plain, labelsep=newline, font={sc, footnotesize}, justification=centering}
    	\caption{Varied Circuit and Control Parameters of Experimental Setup} 
             %\vspace{-.15cm}
    	\label{tab:exp_setup}
        \begin{tabular}{llc}
        \hline \hline
        Symbol            & Description                            & Value                          \\ \hline
        $V_{\mathrm{dc}}$ & DC-link voltage                        & 320 V                          \\
        $f_{\mathrm{sw}}$ & Switching frequency                    & 20 kHz                         \\
        $V_{ll,1}$        & Three-phase voltage base               & 125 V   (1.0 p.u.)             \\
        $S_{1}$           & Inverter VA base                       & 1 kVA   (1.0 p.u.)             \\
        $L_g$             & Grid inductance, nominal/strong        & 15 / 2 mH   (0.3 / 0.04 p.u.)  \\
        $R_g$             & Real circuit ESR, nominal/strong       & 0.533 / 0.216 $\Omega$         \\
        $C_{f}$           & Filter capacitance                   & 4 $\mu$F (0.02 p.u.)           \\
        $\tilde{k}_c$     & Complex current feeding gain           & 1+$j$1.1356                    \\ 
        \hline \hline
        \end{tabular}
        %\vspace{-.5cm}
    \end{table}
    
    \subsubsection{Efficacy of Complex Current Feeding Gain}
    Fig.~\ref{exp_VStability} depicts efficacy of $\tilde{k}_c$ on the base case. The system voltage stability is violated at 0.3~s by a voltage step change, and the increased $\beta_v = 0.95$ is to compensate for the circuit resistance. The unstable pole is $28.49\pm j268.3$~s$^{-1}$ or $42.7$ Hz from the full model, compared to $\lambda_2 = 25.08 -j272.1$ s$^{-1}$ by (\ref{eq:12_}) with 1.5\% error. The unbounded system regains stability right after by setting $\beta_v = -j1.1356$ at 0.4~s without oscillation.
    
        \begin{figure}[!]
            \centering
            \includegraphics[width=1\linewidth]{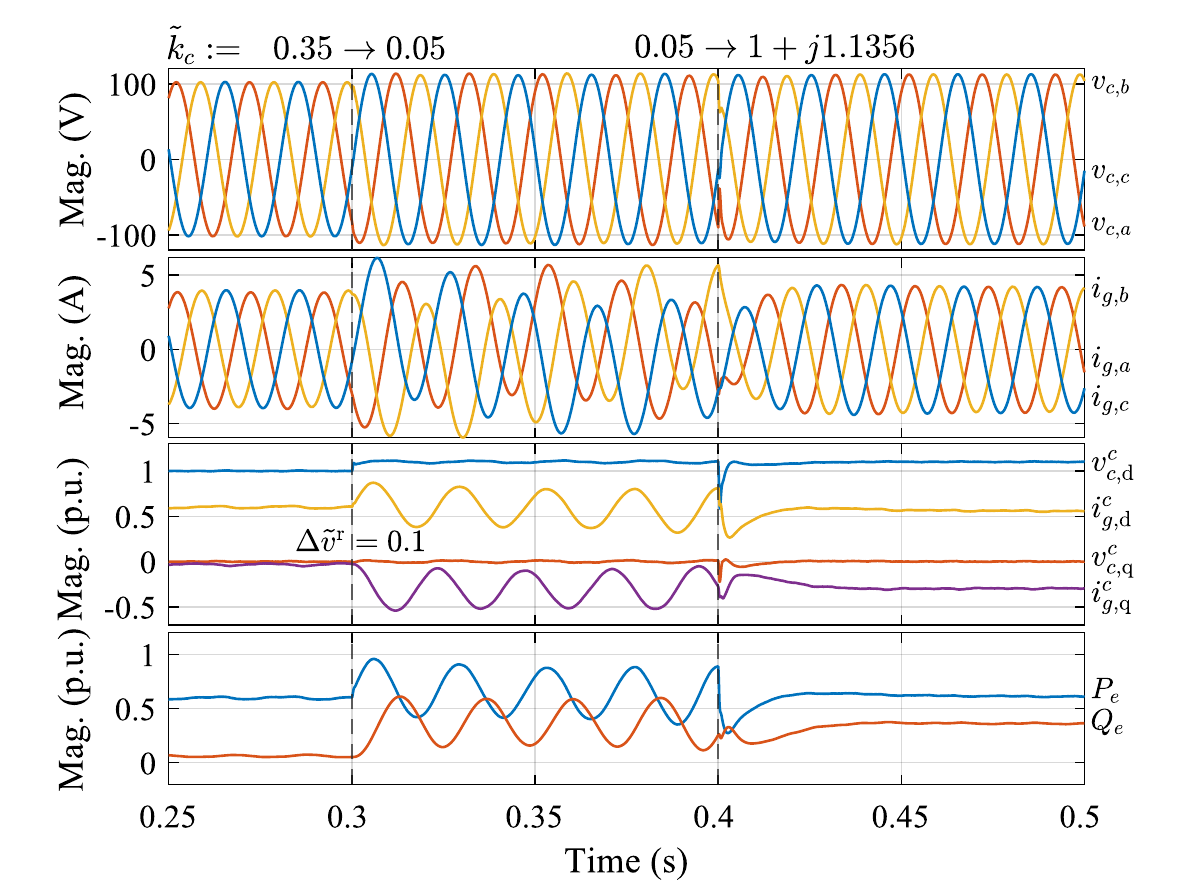}
            \captionsetup{singlelinecheck = false, format= plain, justification=justified, font=footnotesize, labelsep=period}
            \caption{Experiment results of loop stability transition. The base case system transits unstable at 0.3 s by reducing the complex current feeding gain $\tilde{k}_c$ from 0.35 to 0.05 and perturbing a 0.1 p.u. voltage reference change. The system regains stability at 0.4 s by assigning $\tilde{k}_c \coloneqq  1+j1.1356$.
            }
            \label{exp_VStability}
            %\vspace{-.1cm}
        \end{figure}

    \subsubsection{Efficacy of Voltage-angle Compensator}
    Fig.~\ref{exp_VComp_SWING} compares the efficacy of $\mathrm{C}_{v,\theta}$ in suppressing terminal volatility by voltage disturbance with the real power loop as (a) a swing equation or (b) a droop control.
    The perturbed $\Delta \tilde{v}^\mathrm{r}$ triggers power and current peaks during the designed 20~ms voltage transition stage. In both cases, $P_e$ and $I_g$ peaks are reduced by 0.08 p.u. and are constrained under the perturbation magnitude. In addition, $\mathrm{C}_{v,\theta}$ shows its damping functionality, which restrains the noticeable 16\% of voltage overshoot in the droop case.
        \begin{figure}[!]
        %\vspace{-.5cm}
            \centering
            \includegraphics[width=0.98\linewidth]{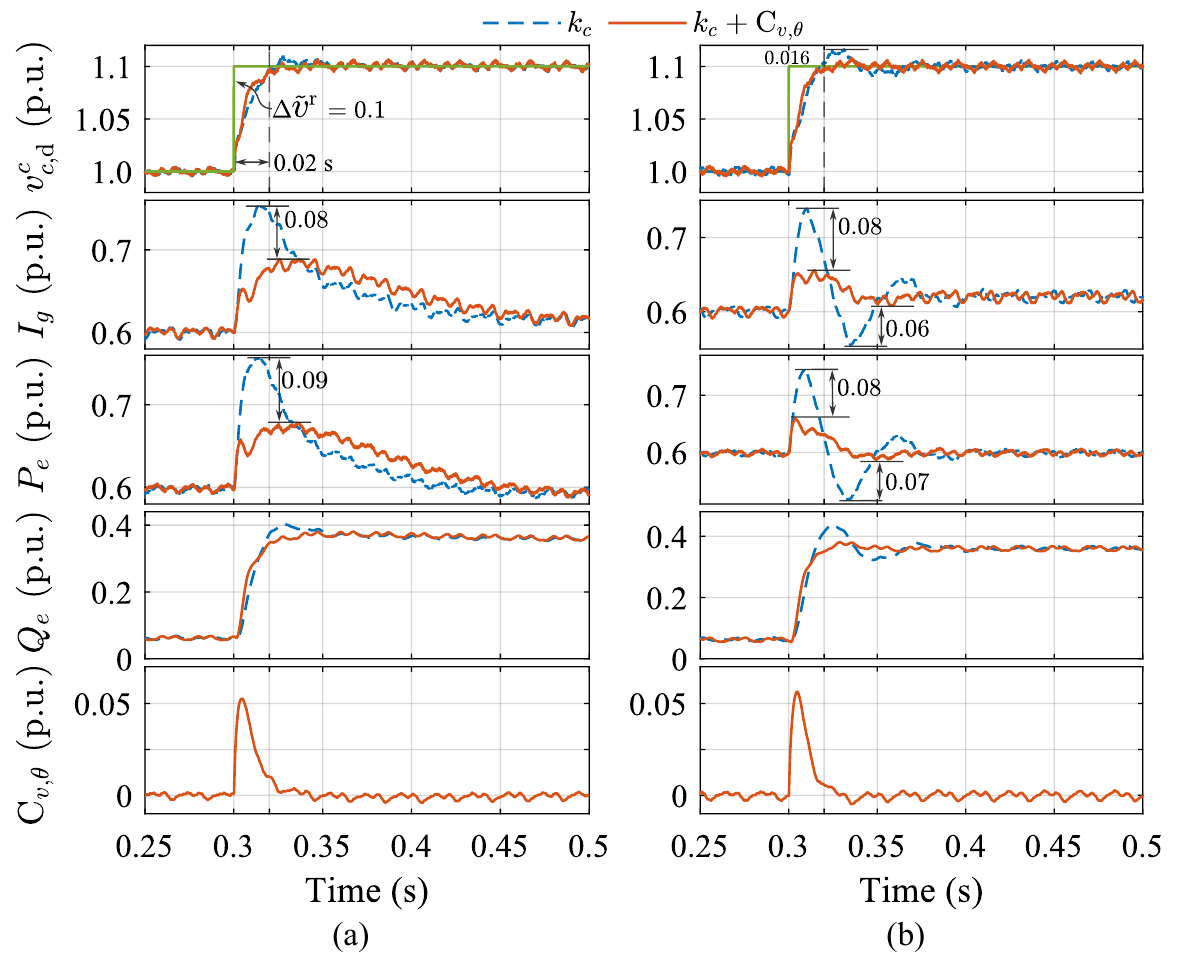}
            \captionsetup{singlelinecheck = false, format= plain, justification=justified, font=footnotesize, labelsep=period}
            \caption{Experiment results comparison of 
             $\mathrm{C}_{v,\theta}$ efficacy of terminal responses, $v^c_{c,\mathrm{d}}$, $I_g$, $P_e$, $Q_e$, $\mathrm{C}_{v,\theta}$, by perturbing $\Delta \tilde{v}^{\mathrm{r}} = 0.1$ p.u. at 0.3 s with two $\mathrm{C}_P$: (a) swing equation $\mathrm{C}_P = \omega_1/(2s^2+66.67s) $, (b) droop $\mathrm{C}_P = 0.1 \omega_1 $. 
            }
            \label{exp_VComp_SWING}
            %\vspace{-.5cm}
        \end{figure}
        
    \subsubsection{Robustness of the Design}
    To validate the robustness of the designed control, two tests are run. Firstly, the inverter is coupled with an ultra-strong grid and performs voltage and power reference changes. It compares the system responses with the original and tuned voltage loop. Secondly, the system faces a sudden phase angle jump. To avoid instability in strong grid conditions, the power droop $\mathrm{C}_P = 0.05 \omega_1 $ is used. Terminal and grid voltages are dropped to 0.98 p.u. to avoid excessive reactive power.

    Fig.~\ref{fig:exp_Xg004} shows terminal responses to (a) 0.02 p.u. voltage reference change and (b) 0.5 p.u. power reference change in a strong grid, $X_g = 0.04$. The voltage loops are tuned at nominal $X_g=0.3$. The refined faster voltage loop has increased inner loop gains that $\{k_{vi}, k_{ip}\}=\{1200,0.80\}$, whereas the $\mathrm{C}_{v, \theta}$ is disabled. In case (a), the voltage response converges in 50 ms where the transition speed is limited by the small $X_g$. $\mathrm{C}_{v,\theta}$ shows its damping again in stabilizing power-caused oscillation, where the damping of the mode around 20 Hz increases from 0.09 to 0.166. A faster voltage loop shows a smoother response on $P_e$. In case (b), the re-tuned voltage loop restrains power oscillation faster than the slower loop does. However, the impact of $\mathrm{C}_{v,\theta}$ is negligible, as the voltage disturbance is insignificant. The power-caused oscillation mode is 11.7 Hz with $\zeta = 0.166$ for the base case, and the mode of the refined voltage loop is 14.5 Hz with $\zeta = 0.2$. In both cases, the faster voltage loop without $\mathrm{C}_{v,\theta}$ provides comparable damping to the slower one with $\mathrm{C}_{v,\theta}$, hence more suitable for a strong grid.
        \begin{figure}[!]
            \centering
            \includegraphics[width=0.98\linewidth]{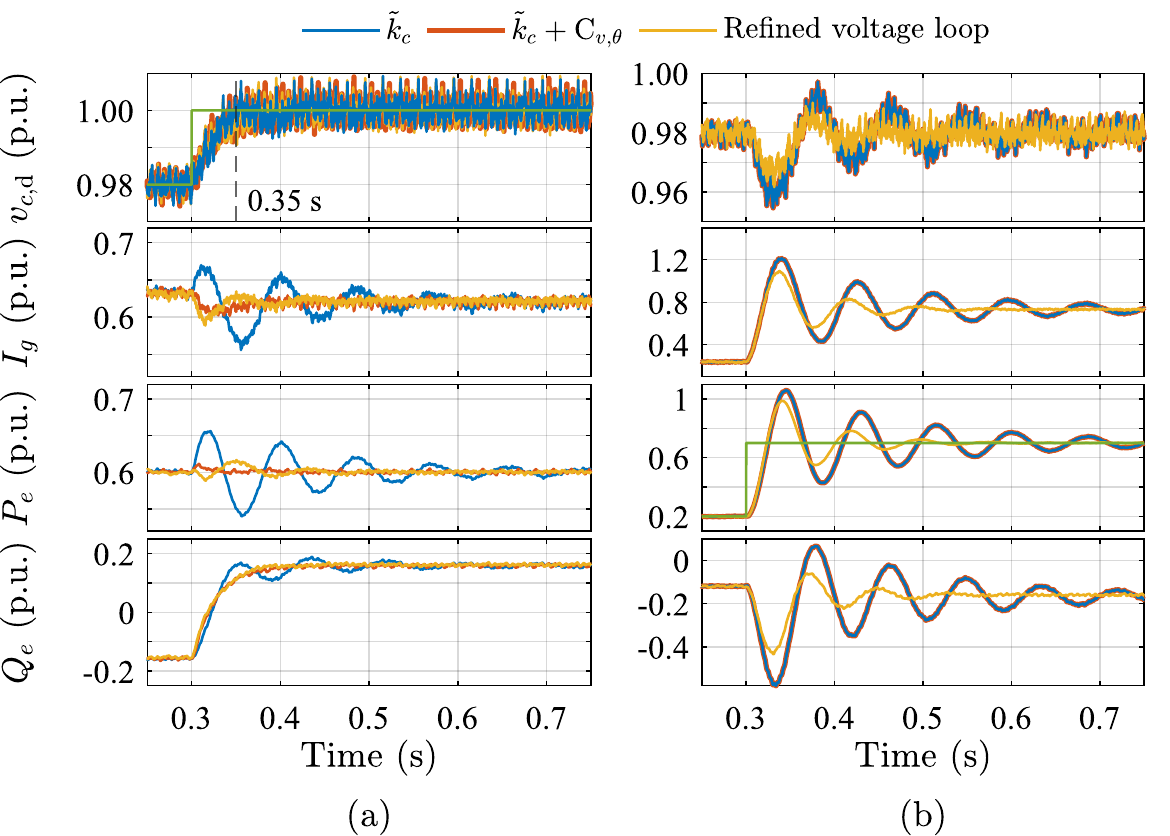}
            \captionsetup{singlelinecheck = false, format= plain, justification=justified, font=footnotesize, labelsep=period}
            \caption{Experimental terminal responses of a VSG coupled with a $X_g = 0.04$ grid to a reference step change of (a) $\Delta \tilde{v}^{\mathrm{r}}$=0.02 and (b) $\Delta P^\mathrm{r}$=0.5. The VSG is aided with (i) complex current feeding gain $\tilde{k}_c$ (ii) with additional voltage-angle compensator $\mathrm{C}_{v,\theta}$ and (iii) refined voltage loop $\{k_{vi},k_{ip}\} = \{1200,0.80\}$.}
            \label{fig:exp_Xg004}
            %\vspace{-.5cm}
        \end{figure}

    Fig.~\ref{fig:exp_PhaseJump60} shows the responses of VSG with designed control facing a sudden phase jump by (a) $60^{\circ}$ change in grid voltage angle at $X_g =0.6$, and (b) change in grid reactance $X_g = 0.04 \mysslash 0.9 \rightarrow 0.9$. In both cases, the system returned stability after the contingency in 20~ms. In case (a), the grid phase jump causes a voltage peak of 1.19 p.u. at $\Delta t = 9$ ms. In case (b), the sudden reactance change causes a severe 1.5 p.u. voltage spike at $\Delta t = 1.55$ ms due to the electromagnetic transient. Though not generalizable, the risk of voltage spikes and insulation stress are to be checked against standards and regulations.
        \begin{figure}[!]
            \centering
            \includegraphics[width=0.98\linewidth]{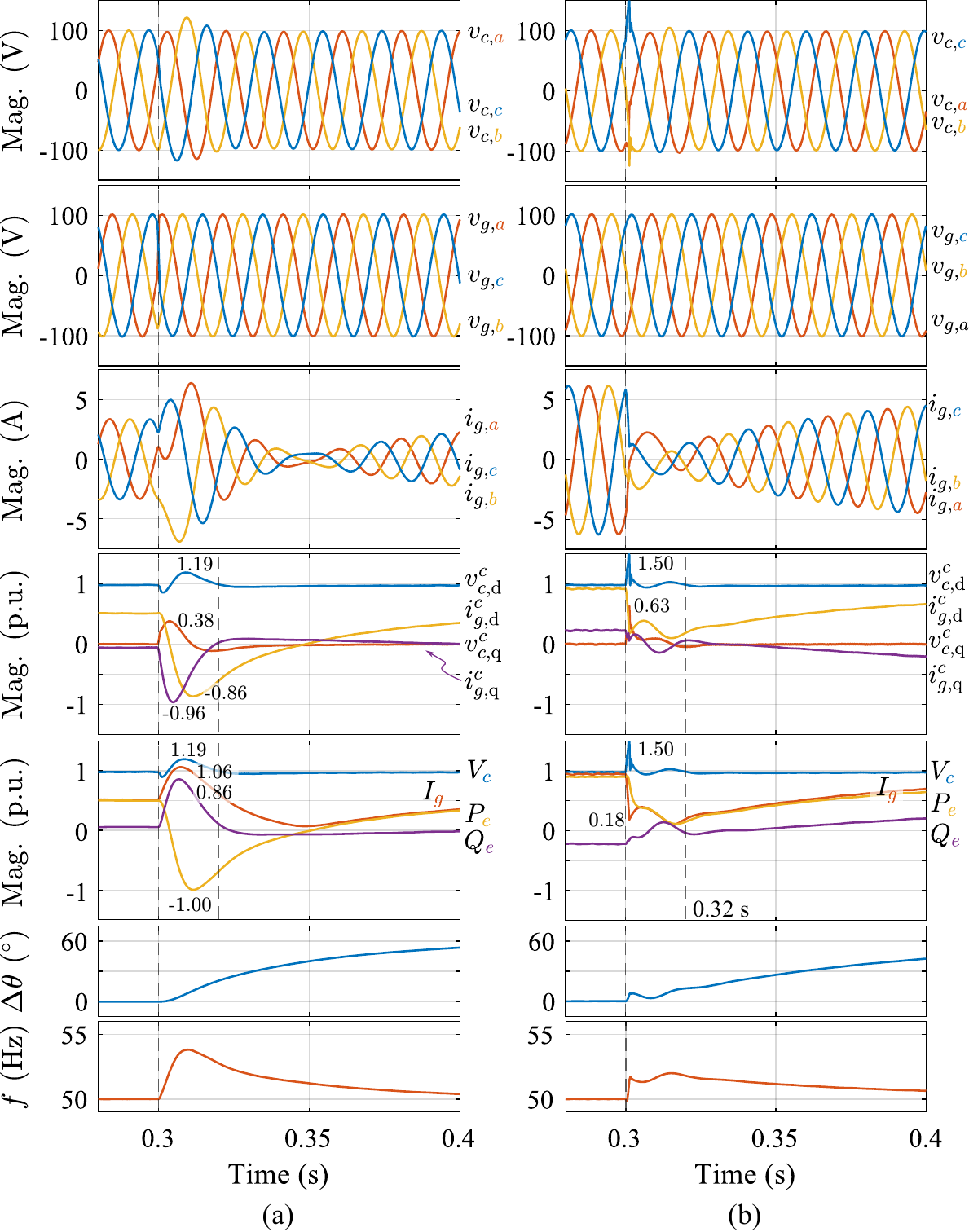}
            \captionsetup{singlelinecheck = false, format= plain, justification=justified, font=footnotesize, labelsep=period}
            \caption{Experimental terminal responses of the proposed VSG control facing a $60^{\circ}$ phase jump by (a) grid voltage angle transition with coupled $X_g=0.6$, and (b) grid reactance transition $X_g = 0.04\,//\,0.9 \rightarrow 0.9$, at 0.3 s.}
            \label{fig:exp_PhaseJump60}
            %\vspace{-.5cm}
        \end{figure}

    {Fig.~\ref{fig:exp_freqDev} depicts the responses of VSG with the designed control facing a sudden grid frequency drop of 1\%. Active power is boosted by 0.67 p.u. according to $D=66.67$. The power and frequency show the closed-loop power swing equation responses with the pole pair at $-16.67\pm15.51$~s$^{-1}$, reaching the peaks after 0.2 s \cite{kundur2022power}. The voltage is intact during the transition.}
        \begin{figure}[!]
            \centering
            % \colorbox{yellow}
            {\includegraphics[width=0.98\linewidth]{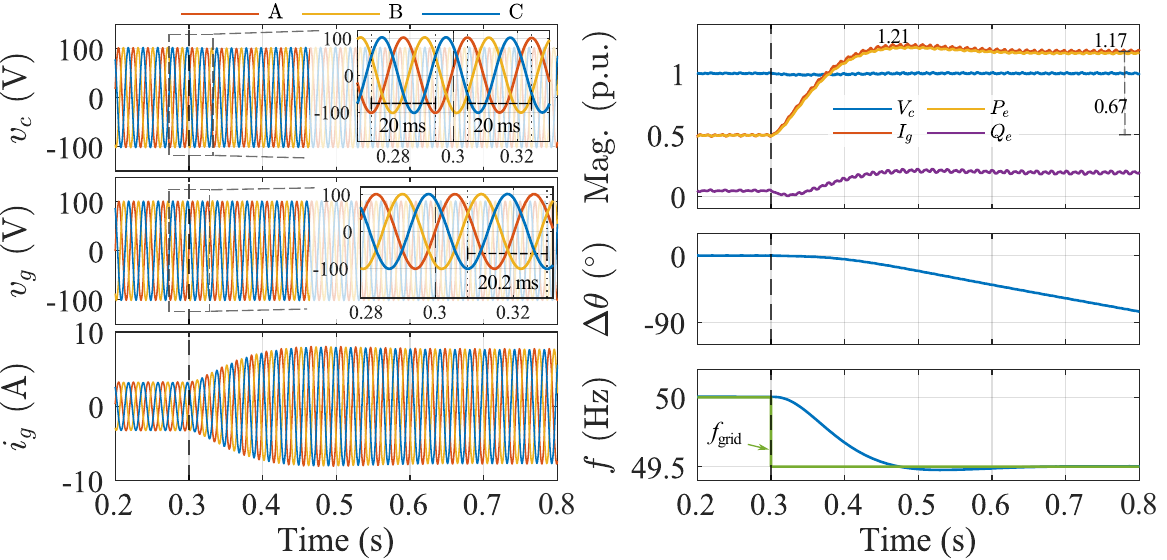}}
            \captionsetup{singlelinecheck = false, format= plain, justification=justified, font=footnotesize, labelsep=period}
            \caption{{Experimental terminal responses of the proposed VSG control facing a 1\% grid frequency drop 0.3~s.}}
            \label{fig:exp_freqDev}
            %\vspace{-.5cm}
        \end{figure}

    These cases demonstrate the robustness of the control design against the wide range of changes in grid reactance (from 0.04 to 0.9~p.u.) and severe disturbances. The voltage loop can restrain terminal voltage fluctuation with proper damping even if control parameters are designed at the nominal grid condition.

    % \subsubsection{Summary}
    \par {
    In summary, the proposed voltage enhancement and induced power decoupling strategies are validated by simulation and experiments. The complex current feeding gain demonstrates its dynamic improvement and robustness against grid transient events. The voltage-angle compensator illustrates its extra damping, helping VSG couple with strong grids. 
    }
    %\vspace{-.3cm}
         
\section{Conclusion} \label{sec:5}
    This paper proposes a solution to an extra-fast and well-damped voltage regulation strategy with straightforward implementation. It largely mitigates the control-interacted sub-synchronous resonance of a VSG,
    ensuring an optimized terminal performance at nominal grid conditions and maintaining robustness against vast grid uncertainties.
    
    The validity is theoretically proven by the proposed geometrical technique that directly analyzes the pole-damping sensitivity of the complex SISO voltage loop.
    The designed voltage-angle compensator suppresses the control impact due to $dq\leftrightarrow3 \phi$ transformation and therefore effectively restrains the transient voltage-induced power fluctuation. 
    The implementation has minimized modification on a generic VSG control scheme. Besides, the parameter tuning procedure is simple by direct pole placement where the nominal grid impedance is the only external input. Due to the ensured robustness, the precise tuning of the voltage loop is not mandatory+. The provided time-domain voltage response estimator simplifies the design workflow.
    
    The time- and frequency-domain responses and robustness are comprehensively validated by theoretical derivation and experiments. In addition to proven performance and stability, a rigorously regulated terminal voltage demonstrates enhanced robustness against severe external disturbances, showing the potential to enhance the transient stability of VSGs. Despite these advantages, the direct angle manipulation brought about by the voltage-angle compensator needs further evaluation of the stability in extreme conditions.
    %\vspace{-.5cm}

\appendix \label{appdx}
    The system is formulated with complex variables \cite{harnefors2007modeling} marked by a tilde. The superscripts `$c$' or `$s$' mean variables in inverter's or grid's $dq$-frame, respectively, whereas the superscript `r' is a variable reference. The subscript `d' or `q' indicates the orthogonal components of the complex variable. A VSG control follows
    \begin{align} 
        \label{eq:3-}
        \Delta \tilde{v}_{s}^{c}=\mathrm{C}_{i}\left( \mathrm{C}_{v}\Delta \tilde{v}^{\mathrm{r}}-\mathrm{F}_{v}\Delta \tilde{v}_c^c+\beta _v\Delta \tilde{i}_{g}^{c} \right) -\mathrm{F}_{i}\Delta \tilde{i}_s^c, 
    \end{align}   
    where $\tilde{v}_{s}^{c}$, $\tilde{v}_{c}^{c}$, $\tilde{i}_{s}^{c}$ and $\tilde{i}_{g}^{c}$ are terminal and PoC voltages, filter and grid currents, respectively. $\tilde{v}^{\mathrm{r}}$ is the voltage set-point. $\Delta$ means a small perturbation.
    $\mathrm{C}_{v}=k_{vp}+k_{vi}/s$ and $\mathrm{C}_{i}=k_{ip}+k_{ii}/s$ are the voltage and current Proportional-integral (PI) controls, respectively.
    $\mathrm{F}_{v}=\mathrm{C}_{v}-{j}B_f, $ and $ \mathrm{F}_{i}=\beta _k\mathrm{C}_{i}-{j}X_f $ are voltage and current feedback loops with decoupling terms ${j}B_C$ and ${j}X_f$, respectively. $\beta_v$ and $\beta_k$ are feeding gains of grid and filter currents, respectively.
    \par
    The adjunct terminal $LC$ filter has grid dynamics as
    \begin{align}
        \label{eq:7-}
        \Delta \tilde{v}_{c}^{s}&=({G}_{L_s}{G}_{C_f}+1)^{-1}\left( \Delta \tilde{v}_{s}^{s}-{G}_{L_s}\Delta \tilde{i}_{g}^{s} \right) ,\\
        \label{eq:8-}
        \Delta \tilde{i}_{s}^{s}&={G}_{C_f}({G}_{L_s}{G}_{C_f}+1)^{-1} ( \Delta \tilde{v}_{s}^{s}+{G}_{C_f}^{-1}\Delta \tilde{i}_{g}^{s} ) ,
    \end{align}
    where ${G}_{L_s}=sL_s+{j}X_s$ and ${G}_{C_f}=sC_f+{j}B_C$ are the filter inductor and capacitor dynamics, respectively, with negligible serial resistance or parallel conductance.
    Further, the $RL$ grid with its impedance $\tilde{Z}_g$ and admittance $\tilde{Y}_g$ is modeled as $\tilde{Z}_g = \tilde{Y}_g^{-1} = sL_g + R_g + {j}X_g$. The current perturbation by $\tilde{v}_{c}^{s}$ and network voltage $\tilde{v}_{n}^{s}$ is
    \begin{align}
        \label{eq:7a}
        \Delta \tilde{i}_g=\tilde{Y}_g\left( \Delta \tilde{v}_{c}^{s}-\Delta \tilde{v}_{n}^{s} \right).
    \end{align}
    Linking the inverter and grid synchronous frames yields vector transformation from inverter's, $\tilde{x}^c$, to grid's $\tilde{x}^s$ by $\Delta \tilde{x}^s=\Delta \tilde{x}^c+j\tilde{x}_0\Delta \delta$, where $\tilde{x}_0$ is the steady-state vector and $\Delta \delta$ is the control imposed angle perturbation.
    \par Connecting the controller (\ref{eq:3-}), filter (\ref{eq:7-},\ref{eq:8-}), and grid (\ref{eq:7a}) by angle dynamics yields the closed-loop VSG system through the input $\tilde{i}_{g}^{s}$ and output $\tilde{v}_{c}^{s}$ being
    \begin{align}
        \nonumber
        T_g\Delta \tilde{v}_{c}^{s}&=\mathrm{C}_i\mathrm{C}_v\left( \Delta \tilde{v}^{\mathrm{r}}-\Delta \tilde{v}_{c}^{s} \right) +{j}G_{\theta ,v}\Delta \theta \\
        \label{eq:5}
        & +\left( sL_s+\beta _k\mathrm{C}_i +jX_{\Delta} \right) \tilde{Y}_g\Delta \tilde{v}_{n}^{s},
    \end{align}
    where 
        \begin{align}
            \label{eq:6}
            G_{\theta ,v}=&\left[ \left( {\mathrm{C}_i}^{-1}+\mathrm{C}_v \right) \tilde{v}_{c}^{0}+k_c\tilde{i}_{g}^{0} \right] \mathrm{C}_i,\\
            \nonumber
            T_g=&1+\left( sL_s+k_c\mathrm{C}_i +jX_{\Delta} \right) \tilde{Y}_g \\
            \label{eq:6-1}
            &+\left( sL_s+\beta _k\mathrm{C}_i +jX_{\Delta} \right) G_{C_f}
        \end{align}   
    With conditions $X_{\Delta} =X_s - X_f =0$ and $B_C = B_f \approx 0$, (\ref{eq:6-1}) can be approximated as 
      \begin{align}
            \label{eq:7}
            T_g            \approx 1+\left( sL_s+k_c\mathrm{C}_i \right) \tilde{Y}_g,
        \end{align}    
    where $\left( sL_s+k_c\mathrm{C}_i \right)$ is the composed current controller;
    $k_c = \beta _k-\beta _v $ is the compound current feeding gain. 
    Approximated $ T_g$ holds since the resonance peak of $LC$ filter is $\omega_1/ \sqrt{X_s B_C}$~rad/s, or hundreds of Hertz, which is beyond the interests of this study. Refs.~\cite{zhang2010modeling,zhao2022robust} show similar approximations that neglect the capacitor dynamics. Similarly, the fast switching delay is neglected.
    
    The primary real and reactive power loops are modeled as
    \begin{align}
        \label{eq:8}
        \Delta \theta &= \mathrm{C}_P (\Delta P^{\mathrm{r}} - \Delta P_e ) , \\
        \label{eq:9}
        \Delta {v}^{\mathrm{r}}_{\mathrm{d}} &= E^{\mathrm{r}} + \mathrm{C}_Q (\Delta Q^{\mathrm{r}} - \Delta Q_e ), 
    \end{align}
    where $\mathrm{C}_P = \omega_B /(2Hs^2+Ds)$ is a swing equation, and $\mathrm{C}_Q = k_q \omega_{Q}/(s+\omega_{Q})$ is a droop controller.
    Substituting $\Delta \theta $ and $\Delta \tilde{v}^{\mathrm{r}}$ by (\ref{eq:8}) and (\ref{eq:9}) with $v_{\mathrm{q}}^{\mathrm{r}}=0$ in (\ref{eq:5}) yields the full VSG control dynamics. Thus, the closed-loop small-signal model of a VSG in an SMIB case is sketched.

    % \begin{align}
    % \left( s-\lambda _1 \right) \left( s-\lambda _2 \right) &=a_2s^2+\tilde{a}_1s+\tilde{a}_0, \\
    % \lambda _1 + \lambda _2 &= - \tilde{a}_1, \\
    % \lambda _1  \lambda _2 &=  \tilde{a}_2. 
    % \end{align}
    % where $ \tilde{a}_1 $ and $ \tilde{a}_2 $ are complex numbers. Throughout my study, $\mathrm{arg}(\tilde{a}_0)=90^{\circ}$ is valid. And I can place $ \tilde{a}_1 $ anywhere on a complex plane. So the most sensible and convenient way is just to put two poles on $-45^{\circ}$. Therefore, $\mathrm{arg}(\lambda_1\lambda_2)=\mathrm{arg}(-\lambda_1)+\mathrm{arg}(-\lambda_2)=\mathrm{arg}(\tilde{a}_0)=90^{\circ}$, and since $\lambda_{1,2}$ are of the same angle,   $\mathrm{arg}(\lambda_1+\lambda_2)=\mathrm{arg}(\lambda_1)=\mathrm{arg}(\lambda_2)=\mathrm{arg}(\tilde{a}_1)=-45^{\circ}$.


\end{document}